\documentclass{article}[12pt]
\usepackage{amsmath,amssymb,amsfonts}
\usepackage{graphics, subfigure,color} 

\usepackage{latexsym}
\usepackage[dvips]{graphicx}
\usepackage{epsfig}
\usepackage{braket}

\usepackage{hyperref}

\numberwithin{equation}{section}

\newtheorem{prop}{Proposition}[section]

\newtheorem{lemma}[prop]{Lemma}

\newtheorem{theorem}[prop]{Theorem}

\newcommand{\cE}{{\cal{E}}}

\newcommand{\cF}{{\cal{F}}}
\newcommand{\cV}{{\cal{V}}}
\newcommand{\cB}{{\cal B}}
\newcommand{\cT}{{\cal T}}
\newcommand{\cS}{{\cal S}}
\newcommand{\cJ}{{\cal J}}
\newcommand{\cL}{{\cal L}}
\newcommand{\cH}{{\cal H}}
\newcommand{\cM}{{\cal M}}
\newcommand{\cD}{{\cal D}}
\newcommand{\cC}{{\cal C}}
\newcommand{\bQ}{{\bf Q}}
\newcommand{\bV}{{\bf V}}
\newcommand{\bX}{{\bf X}}
\newcommand{\bA}{{\bf A}}
\newcommand{\cQ}{{\cal{Q}}}
  
\newcommand{\Tr}[1]{\:{\rm Tr}\,#1}
 
\newcommand{\bbone}{{\bf 1}}

\newcommand{\bee}{\begin{equation}}
\newcommand{\ee}{\end{equation}}
\newcommand{\beq}{\begin{eqnarray}}
\newcommand{\eeq}{\end{eqnarray}}
\newcommand{\bqa}{\begin{eqnarray}}
\newcommand{\eqa}{\end{eqnarray}}
\newcommand{\bea}{\begin{eqnarray}}
\newcommand{\eea}{\end{eqnarray}}
\newcommand{\beann}{\begin{eqnarray*}}
\newcommand{\eeann}{\end{eqnarray*}}  
  
\newcommand{\tr}{{\rm Tr}}
\newcommand{\prf}{{\noindent \bf Proof\; \; }}
\newcommand{\qed}{{\hfill $\Box$}}

\makeatletter

\begin{document}

\title{Constructive Tensor Field Theory: The $T^4_3$ Model}

\author{
Thibault Delepouve\footnote{delepouve@cpht.polytechnique.fr, Laboratoire de Physique Th\'eorique, CNRS UMR 8627, Universit\'e Paris Sud, 91405 Orsay Cedex, France
 and Centre de Physique Th\'eorique, CNRS UMR 7644, \'Ecole Polytechnique, 91128 Palaiseau Cedex, France.}\;\; and
Vincent Rivasseau\footnote{rivass@th.u-psud.fr, Laboratoire de Physique Th\'eorique, CNRS UMR 8627, Universit\'e Paris Sud, 91405 Orsay Cedex, France
 and Perimeter Institute for Theoretical Physics, 31 Caroline St. N, N2L 2Y5, Waterloo, ON, Canada.}}

\maketitle
\begin{abstract} We build constructively the simplest tensor field theory which requires some renormalization, namely the rank
three tensor theory with quartic interactions and propagator inverse of the Laplacian on $U(1)^3$.
This superrenormalizable tensor field theory has a power counting almost similar to ordinary $\phi^4_2$.
Our construction uses the multiscale loop vertex expansion (MLVE) recently
introduced in the context of an analogous vector model. 
However to prove analyticity and Borel summability of this model requires new
estimates on the intermediate field integration, which is now of matrix rather than of scalar type.
\end{abstract}

\section{Introduction}
\label{intro}

Colored tensor models \cite{Gurau:2009tw,Gurau:2011xp} were proposed first as an 
improvement of group field theory \cite{Boulatov:1992vp,Oriti:2011jm}. A key progress over earlier tensor models \cite{tens1,tens2,tens3}
is that they admit a $1/N$ expansion \cite{Gurau:2010ba,Gurau:2011aq,Gurau:2011xq}.  

$U(N)^{\otimes D}$ invariant actions and observables for a pair of rank $D$ complex conjugate tensor fields of dimension $N$
are in one to one correspondence with $D$-regular  edge-colored bipartite graphs  \cite{Gurau:2011kk}. 
Such interactions generalize the invariant matrix interactions used
in matrix models \cite{Di Francesco:1993nw} and in 
matrix based field theories such as the Kontsevich model \cite{Kontsevich:1992ti}Ê or the renormalizable asymptotically safe non-commutative
Grosse-Wulkenhaar model over the four dimensional Moyal space \cite{Grosse:2004yu,Disertori:2006nq,Rivasseau:2007ab,Grosse:2009pa,Grosse:2012uv}. 
Random tensor models with such invariant actions (called ``uncolored" models \cite{Bonzom:2012hw}) are the effective theories 
coming from colored models when integrating out all tensor fields \emph{save one}. They 
also admit a $1/N$ expansion which is universal in a certain precise mathematical sense \cite{Gurau:2011kk}.

The \emph{tensor track} \cite{Rivasseau:2011hm,Rivasseau:2012yp,Rivasseau:2013uca} is the proposal to use 
the infinite dimensional space of tensor invariant interactions as a new \emph{theory space} \cite{Rivasseau:2014ima} for the quantization
of gravity in dimensions higher than 2. In particular it proposes to
study renormalization group flows in this space \cite{Benedetti:2014qsa}   in the hope to discover interesting new random geometries.
Indeed the Feynman graphs of rank $D$ tensor theory are $(D+1)$-regular edge-colored bipartite graphs dual to triangulations of (pseudo)-manifolds of dimension $D$. 
Conversely \emph{any} $D$-dimensional pseudo-manifold is dual to infinitely many Feynman graphs of a rank $D$ tensor model.
Hence the perturbation expansion of tensor field theories performs a sum over all $D$-dimensional (pseudo)-manifolds. 
Moreover this expansion sums also over discretized metrics. In particular tensor amplitudes without further data
ponder equilateral triangulations exactly with a discretized 
form of the Einstein-Hilbert action \cite{ambjorn}. It should also be noticed that adding group field theoretic projectors to tensor models leads to tensor amplitudes
which are spin foams, achieving second quantization of loop quantum gravity \cite{Oriti:2013aqa}.

To launch and study a renormalization group flow in the tensor theory space  
requires to introduce convenient cutoffs allowing for scale decomposition. 
The most convenient field-theoretic way to do this is to introduce as propagator an inverse Laplacian that softly breaks the $U(N)^{\otimes D}$ invariance
and to use the heat-kernel regularization.
This procedure can be justified also out of perturbative renormalization considerations \cite{Geloun:2011cy}.

Tensor field theories \cite{BenGeloun:2011rc} have been therefore defined as random tensor models with tensor invariant interactions
and such a Laplacian-based propagator. 
The tensorial $1/N$ expansion is the essential tool which allows for power counting and for
renormalization of such theories, just like the matrix $1/N$ expansion does in the Grosse-Wulkenhaar theory  \cite{Grosse:2004yu}. 

Superrenormalizable and renormalizable tensor field theories come essentially in two versions. 
The basic version has no group field theoretic projectors
\cite{BenGeloun:2011rc,BenGeloun:2012pu,Geloun:2013saa} and can be considered the field theoretic 
version of random tensor models. It sums over triangulations simply equipped with the graph-distance metric, hence is a kind of
\emph{equilateral} version of Regge calculus.
The more sophisticated version equipped with additional group-field theoretic projectors 
\cite{Carrozza:2012uv,Carrozza:2013wda,Carrozza} uses a different metric which incorporates the usual 
simplicity constraints of group field theory, hence should be properly called tensor group field theory. In both cases
renormalizable models have the generic property of being
asymptotically free\footnote{The tensor theory space is different 
from the ``Einsteinian" theory space studied in the asymptotic safety program \cite{reuter}, which is the space of diffeo-invariant functions of a metric $g_{\mu \nu}$ 
on a fixed ${\mathbb R}^4$ topology. Therefore there is absolutely no contradiction between existence of a non-Gaussian fixed
point in Einsteinian space and asymptotic freedom in the tensorial space. The uv asymptotically free tensorial flow can lead 
in the infrared to one or presumably several phase transitions which could create a background random space with effective local properties similar to $R^4$. The same flow
rewritten in new effective variables could then look as if it emerges out of the vicinity of an asymptotically safe fixed point on this effective background space.}
\cite{BenGeloun:2012pu,BenGeloun:2012yk,Carrozza:2014rba}. 
This is up to now the physically most interesting result of the tensor track, since it allows to envision 
geometrogenesis \cite{Konopka:2006hu} Êas a cosmological scenario \cite{Gielen:2013kla} of tensor theories.

Constructive field theory \cite{GJ,Riv} is a set of techniques to resum perturbative quantum field theory 
and obtain a rigorous definition of quantities such as the Schwinger functions of interacting renormalizable models.
The loop vertex expansion (LVE) \cite{Rivasseau:2007fr,Magnen:2007uy,Rivasseau:2010ke,Rivasseau:2013ova} 
is a constructive tool well-adapted to the control of \emph{non-local} theories in a single renormalization group slice.
It is also particularly efficient for the non-perturbative construction of 
random tensor models \cite{sefu1,Gurau:2013pca,Delepouve:2014bma}. 
A multiscale loop vertex expansion or MLVE has been recently defined and tested on a vector field theory \cite{MLVE}. 
To include renormalization, this MLVE adds to the usual Bosonic layer of the LVE
a Fermionic layer (Mayer-type expansion \cite{mayer,brydgesfed,brydges1}). It has been used to 
revisit the standard construction of the $\phi^4_2$ theory \cite{Rivasseau:2014bya}.

It is therefore natural to extend the constructive program to tensor field theories.
This is what we do in this paper for the simplest such theory which requires some infinite renormalization, namely the $U(1)$
rank-three model with inverse Laplacian propagator and quartic interactions, which we nickname $T^4_3$. It can
also be considered as an ordinary field theory on the torus $T^3$, but with non-local quartic interactions which break rotation invariance.
 It turns out that the 
$T^4_3$ model requires to add to the MLVE of \cite{MLVE}Ê
several additional non-trivial arguments, since the tensor propagator links the indices 
of the tensor together and the intermediate fields are matrices rather than scalars.

The plan of the paper is the following. In section 2 we recall the model and its intermediate field representation
and we introduce the standard multiscale analysis \cite{Riv,BenGeloun:2011rc} to perform renormalization.

In section 3 we perform the MLVE itself, which expresses
the connected functions of the theory as a two-level tree expansion, with both Bosonic and Fermionic links. 
We also state our main theorem which is the convergence of this expansion, allowing to prove existence of the ultraviolet
limit of the theory and its Borel summability in a certain cardioid-like domain of the coupling constant.

In section 4 we gather the proofs of the theorem. The Fermionic integrals are exactly similar to those of 
\cite{MLVE} and bounded in the same way. We decompose then the Bosonic blocks into perturbative and non perturbative
parts which we evaluate separately thanks to a Cauchy-Schwarz inequality. The non-perturbative part requires to bound
a determinant which is new compared to \cite{MLVE}; this is done through a combination of norms and trace bounds. The perturbative part requires a
parametric representation of resolvents factors which allows strand factorization and resolvent bounds in the style of \cite{Gurau:2013pca}.
Concluded by a relatively standard perturbative bound on convergent graphs with scales constraints, this part delivers the key 
power counting factors which ultimately beat the combinatorics of the expansion
in the same manner than in \cite{MLVE}.

\medskip
\noindent{\bf Acknowledgments}
We thank warmly Razvan Gurau for useful discussions and for sharing with us insights on the multiscale loop vertex expansion, and Fabien Vignes-Tourneret
for pointing out an important correction to the initial version of this paper.
V. Rivasseau also acknowledges the partial support of the Perimeter Institute.

\section{The Model}

\subsection{Laplacian, Bare and Renormalized Action}

We shall use the time-honored constructive practice to write $O(1)$ 
for any inessential numerical constant throughout this paper.

Consider a pair of conjugate rank-3 tensor fields $T_n, \bar T _{\bar n}$ with $n = \{ n_1,n_2,n_3 \} \in \mathbb{Z}^3$, 
and $\bar n = \{\bar n_1,\bar n_2,\bar n_3 \} \in \mathbb{Z}^3$. They belong respectively to the tensor product 
$\cH = \cH_1 \otimes \cH_2 \otimes\cH_3$ and to its dual,
where each $\cH_c$ is an independent copy of  $\ell_2 (\mathbb{Z})= L_2 (U(1))$, and  the color or strand index $c$ takes values $c=1,2,3$.
Indeed by Fourier transform these fields can be considered also as ordinary scalar fields  
$T  (\theta_{1},\theta_{2},\theta_{3})$ and $\bar T(\bar \theta_{1},\bar \theta_{2},\bar \theta_{3} )$  
on the three torus ${\rm \bf T}_3 = U(1)^3$ \cite{BenGeloun:2011rc}.

\bee d\mu_{C} (T, \bar T) =
\biggl( \prod_{n, \bar n \in {\mathbb Z}^3} \frac{dT_n dT_{\bar n}}{2i \pi } \biggr)  
[{\rm Det}
\, C]^{-1} 
e^{-\sum_{n, \bar n}  
T_n  C_{n, \bar n}^{-1} T_{\bar n}   }
\ee
where the bare propagator $C$ for simplicity has unit mass:
\bee C_{ n, \bar n} = \delta_{n, \bar n} C(n) , \quad C(n) \equiv \frac{1}{ n^2_1 + n_2^2 + n_3^2 +1}. \label{propdef}
\ee
The bare partition function is then
\bee  Z_0(g) = \int   e^{-\frac{g}{2}    \sum_c V^c (T, \bar T)}    d\mu_{C} (T, \bar T)
\ee
where $g$ is the coupling constant and 
\bee  V^c(T, \bar T) =\sum_{n,\bar n, p,\bar p} \left( T_n\bar T_{\bar n}  \prod_{c'\neq c}\delta_{n_{c'} \bar n_{c'}}      \right) 
\delta_{n_c \bar p_c} \delta_{p_c \bar n_c} \left(\ T_p\bar T_{\bar p} \prod_{c'\neq c}\delta_{p_{c'} \bar p_{c'}}     \right)  \label{formulavc}
\ee
are the three quartic interaction terms of random tensors at rank three. This model is the \emph{simplest interacting tensor field theory}.
Indeed it has smallest rank (three), smallest interaction degree (quartic), is symmetric under independent unitary 
transforms in each of the three spaces of the tensor product $\cH = \cH_1 \otimes \cH_2 \otimes\cH_3 $ and globally symmetric under color permutations. 
Remark that all three quartic interactions are \emph{melonic} at rank 3. This is no longer true at higher rank \cite{Delepouve:2014bma}. 

\begin{figure}[!h]
  \begin{center}

  {\includegraphics[width=0.55\textwidth]{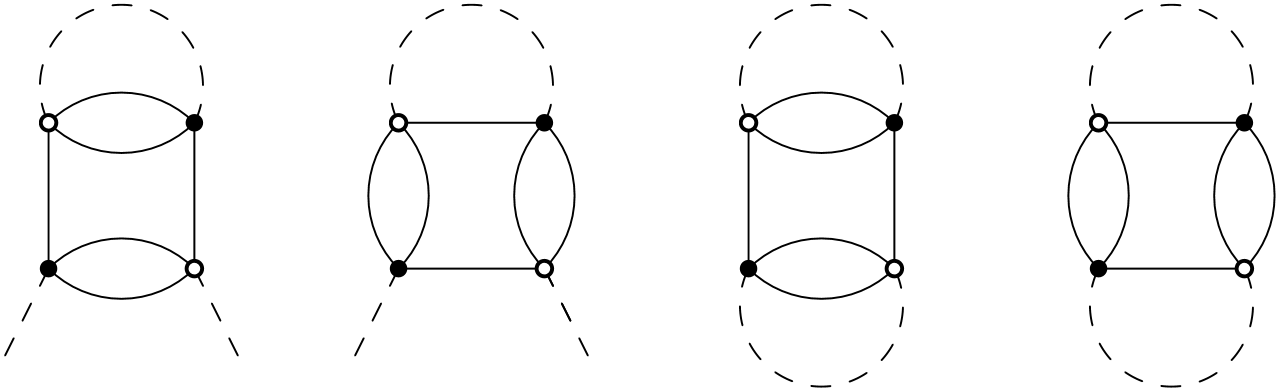}}
   \end{center}
  \caption{From left to right, the divergent self-loop $\cM$, the convergent self loop and the two vacuum connected graphs $\cV_1$ and $\cV_2$.}
  \label{divergences}
\end{figure}

The model has a power counting almost similar to the one of ordinary $\phi^4_2$ \cite{nelson,Simon}. 
It has for each color $c$ two vacuum divergent graph, both with a single vertex; $\cV_1$ is linearly 
divergent and $\cV_2$ is logarithmically divergent. It has also a single logarithmically divergent two-point graph $\cM$, again with a single vertex,
which requires a mass renormalization (see Figure \ref{divergences}).

To have well defined equations and quantities we impose a cutoff $N$, i.e. we replace $\cH_c$ by $\ell_2 ([-N,N])$.
and from now on in this section all sums over indices such as $n, \bar n$ are therefore restricted to belong to $ [-N, N]^3$.

The bare amplitude for $\cM$ is the sum of three amplitudes with color $c$, each of which is a non-trivial function of the single incoming momentum $n_c$
\bee  A(\cM)  = \sum_c   A(\cM_c),  \quad  A(\cM_c) (n_c) =  -g  \sum_{p  \in  [-N, N]^3}  \frac{\delta (p_c -n_c) }{p^2 + 1} .
\ee
The sum over $p$ diverges logarithmically as $N \to \infty$. The mass counterterm is minus the value at $n_c =0$, namely
\bee  \Delta m = -\sum_c A(\cM_c) (0) = g \sum_c  \delta m^c,   \quad  \delta m^c  = \sum_{p  \in  [-N, N]^3}  \frac{\delta(p_c) }{p^2 + 1} = 
\sum_{p  \in  [-N, N]^2}  \frac{1}{p^2 + 1} . \label{defdelta1}
\ee
Remark that $ \delta m^c $ is independent of $c$, so that in fact 
\bee  \Delta m = 3 g  \sum_{p  \in  [-N, N]^2}  \frac{1}{p^2 + 1} .
\ee
The renormalized amplitude of $\cM$ at color $c$ is a convergent sum, hence no longer requires the cutoff $N$:
\bea  A^{ren}(\cM_c) (n_c)&=&  A(\cM_c) (n_c)  + \delta m^c = - g  \sum_{p   \in {\mathbb Z}^3}   \frac{\delta (p_c -n_c)  - \delta (p_c) } 
{p^2+ 1} = g  A(n_c)\\
A(n_c) &=& \sum_{p   \in {\mathbb Z}^2}   \frac{ n_c^2} {(n_c^2 + p^2+ 1)(p^2 +1)}  \le  O(1) \log (1+ \vert n_c \vert) . \label{Abound}
\eea

We should similarly compute the vacuum counterterms,  taking into account the presence of the $\Delta m$ counter term. 
The partition function with this mass counter term included is 
\bee  Z_1(g) = \int   e^{-\frac{g}{2}    \sum_c V^c (T, \bar T) + g \sum_c \delta m^c \sum_n T_n \bar T_n   }    d\mu_{C} (T, \bar T).
\ee

The graph $\cV_1$ is the sum of three colored subgraphs $\cV_1^c$. $\cV^1_1$ requires the counterterm
\bee \delta \cV_1^1 =  \frac{g}{2} \sum_{n_1 , n_2, n_3, p_2, p_3 } \frac1{n_1^2+ n_2^2 +n_3^2 +1} \frac1{n_1^2 + p_2^2 +p_3^2 +1} .
\ee
Similarly the graph $\cV_2$ is the sum of three colored subgraphs $\cV_2^c$. $\cV_2^1$ requires the counter term
\bee \delta \cV_2^1    =  \frac{g}{2}\sum_{n_1 , n_2, n_3, p_1 } \frac1{n_1^2+n_2^2+n_3^2 +1} \frac1{p_1^2+n_2^2+n_3^2 +1} .
\ee
Finally the mass counterterm itself generates a divergent vacuum graph $\cV_{\Delta m}$ which requires a different counterterm:
\bee \delta\cV_{\delta m}^1 = - g  \sum_{n_1 , n_2, n_3, p_2, p_3 } \frac1{n_1^2+ n_2^2 +n_3^2 +1} \frac1{p_2^2 + p_3^2 +1}.
\ee
The counterterms for colors 2 and 3 are obtained by 
the same formulas with the appropriate color permutation.
The renormalized partition function is therefore
\bee    Z(g) = e^{ \sum_c (\delta \cV_1^c + \delta \cV_2^c + \delta\cV_{\delta m}^c) } 
\int   e^{-\frac{g}{2} \sum_c V^c (T, \bar T) }e^{ g \sum_c \delta m^c \sum_n T_n \bar T_n  }  d\mu_{C} (T, \bar T) .
\ee 

Such quartic tensor models are best studied in the intermediate field representation \cite{Gurau:2013pca}. 
We put $g = \lambda^2$ and decompose the three interactions $V^c$ in \eqref{formulavc} by introducing 
three intermediate Hermitian matrix  fields $\sigma^c$ acting on $\cH_c$, in the following way
\bee
 e^{-\frac{\lambda^2}{2} V^c(T, \bar T) }= \int e^{i\lambda \sum_{n,  \bar n} \prod_{c'\neq c}\delta_{n_{c'} \bar n_{c'}} 
 \left( T_n\bar T_{\bar n}   \right)\sigma^c_{n_c\bar n_c}}    d\nu(\sigma^c).
\ee
where $d\nu(\sigma^c)$ is the normalized Gaussian independently identically distributed measure of covariance 1 on 
each independent coefficient of the Hermitian matrix  $\sigma^c$. It is convenient to consider $C$ as a (diagonal) operator acting on
$\cH =\cH_1 \otimes \cH_2 \otimes\cH_3 $, and to 
define in this space the operator
\bee {\vec \sigma }= \sigma^1\otimes\mathbb{I}_2\otimes\mathbb{I}_3 +\mathbb{I}_1\otimes\sigma^2\otimes\mathbb{I}_3 
+\mathbb{I}_1\otimes\mathbb{I}_2\otimes\sigma^3
\ee 
where $\mathbb{I}_c$ is the identity over $\cH_c$.

We can absorb the mass counterterm in a translation of the quartic interaction (in a way somewhat analogous to Wick-ordering). Indeed remark that
\bee 
\frac{1}{2}  V^c (T, \bar T)  -  \delta m^c \sum_n T_n \bar T_n  +\frac{1}{2}    \sum_{n, p} \delta_{n_c , p_c} \frac{  1 }{ (n^2 -n_c^2 )+ 1} \frac{  1 }{ (p^2 -p_c^2 )+ 1} 
=    \label{reninter}
\ee
\bee
\sum_{n,\bar n, p,\bar p}  \biggl[  \prod_{c'\neq c}\delta_{n_{c'} \bar n_{c'}} \delta_{p_{c'} \bar p_{c'}} \left( T_n\bar T_{\bar n} -   
 \frac{   \delta_{n_c \bar n_c}  }{ (n^2 -n_c^2 )+ 1}    \right)   \biggr]
\delta_{n_c \bar p_c} \delta_{p_c \bar n_c} 
 \biggl[   \prod_{c'\neq c} \delta_{p_{c'} \bar p_{c'}}  \left(\ T_p\bar T_{\bar p} - \frac{ \delta_{p_c \bar p_c} }{(p^2 -p_c^2)+ 1}   \right)  \biggr]  . \nonumber
\ee
Therefore, defining
\bee \delta \cV_3^1    =  \frac{\lambda^2}{2}\sum_{n_1 , n_2, n_3, p_2, p_3  \in  [-N, N]^5 } \frac1{n_2^2+n_3^2 +1} \frac1{p_2^2+p_3^2 +1} = \frac{\lambda^2}{2} N (\delta m^c)^2 ,
\ee
$Z(g)$ can be evaluated from \eqref{reninter} as
\begin{eqnarray}
Z(g)&=&  e^{ \sum_c (\delta \cV_1^c + \delta \cV_2^c + \delta \cV_3^c + \delta\cV_{\delta m}^c) }  \int d\nu(\vec \sigma)
  d\mu_{C}(T, \bar T)  e^{ i\lambda  \sum_c \left( \sum_{n, \bar n}T_n{\bar T}_{\bar n}
 \sigma^c_{n_c \bar n_c}\prod_{c'\neq c}\delta_{n_{c'} \bar n_{c'}}-  \delta m^c {\tr_c }\sigma^c   \right)   }   \nonumber  \\
 &=&   Z'(g)  \int d\nu(\vec \sigma)  \prod_c  e^{-i\lambda  \sum_c \delta m^c {\tr_c }\sigma^c   } e^{ 
 -{\rm \bf Tr} \log \left[\mathbb{I}-i\lambda    C^{1/2}  \vec \sigma C^{1/2}  \right]  }  .
 \label{log2}
\end{eqnarray}
In this equation, $\tr_c$ means a trace over $\cH_c$, ${\rm \bf Tr}$ means trace on the tensor product 
$\cH = \cH_1 \otimes \cH_2 \otimes\cH_3 $, $\mathbb{I}$ is the identity on $\cH$, 
$d\nu(\vec \sigma)= \prod_c  d\nu(\sigma^c)$,
$R$ is the \emph{resolvent} operator on $\cH$
\bee 
R (\vec \sigma)  \equiv \frac{1}{  \mathbb{I} -i \lambda C^{1/2}  \vec \sigma C^{1/2}  } ,
\ee
and
\bee   Z' (g) =  e^{ \delta \cV_1^c + \delta \cV_2^c + \delta \cV_3^c + \delta\cV_{\delta m}^c} .
\ee

We remark that the first term in the expansion in $\lambda$ of $-{\rm \bf Tr} \log \left[\mathbb{I}-i\lambda C^{1/2}  \vec \sigma C^{1/2}  \right] $ combines nicely with the term
$-i\lambda  \sum_c \delta m^c {\tr_c }\sigma^c  $, since
\bee {\rm \bf Tr} \, C {\vec \sigma } - \sum_c \delta m^c {\tr_c }\sigma^c  = \sum_c  \sum_{n_c}A(n_c) \sigma^c_{n_c n_c}. \label{vecd}
\ee

Joining \eqref{log2} and \eqref{vecd}Ê
gives
\bee  \label{niceequat1}
Z(g) = Z' (g)  \int d\nu(\vec \sigma)  e^{ i \lambda \sum_c  \sum_{n_c}A(n_c) \sigma^c_{n_c n_c} -{\rm \bf Tr} \log_2 \left[\mathbb{I}-i\lambda C^{1/2}  \vec \sigma C^{1/2}  \right]    } ,
\ee
where $\log_2 (1-x) \equiv x+ \log (1-x) = O(x^2)$.

Finally we should rework $Z' (g)$ to check that it compensates indeed the divergent vacuum graphs of the $\sigma$ functional integral.
First remark that $ \delta \cV_1^c  + \delta \cV_3^c + \delta \cV_{\delta m}^c$ nicely recombine as
\bee \cD  \equiv \sum_c \delta \cV_1^c  + \delta \cV_3^c + \delta \cV_{\delta m}^c  = \frac{\lambda^2}{2}  \sum_c \sum_{n_c}  A^2(n_c)
\ee
where for the last equality we recall that $\sigma^c$ cannot contract to $\sigma^{c'}$ for $c' \ne c$.
Similarly we remark that the non-melonic 
log-divergent counter term for $\cV_2$ can be written as a $\sigma$ integral namely
\bee \cE= \sum_c \delta \cV^c_2 = \frac{\lambda^2}{2}  \int d\nu(\vec \sigma)  {\rm \bf Tr}  ( C {\vec \sigma })^2 .
\ee
Therefore
\bee  \label{niceequat2}
Z(g) = \int d\nu(\vec \sigma) e^{- {\rm \bf Tr} \log_2 \left[\mathbb{I}-i\lambda C^{1/2}  \vec \sigma C^{1/2}  \right] +  i \lambda  \sum_c  \sum_{n_c}A(n_c) \sigma^c_{n_c n_c}  + \cD + \cE  } .
\ee
As expected, thanks to $\cD$ and $\cE$, the first order term in $\lambda^2$ cancels exactly in this $\sigma$ representation of $\log Z(g)$, as they did
in the $(T, \bar T )$ representation, so that
\bee  \log Z(g)  = O(\lambda^4)  = O(g^2).
\ee

We remark now that 
\bee 
d\nu(\vec \sigma) e^{ i \lambda  \sum_c  \sum_{n_c}A(n_c) \sigma^c_{n_c n_c} + \cD } 
\ee
is exactly the Gaussian normalized measure for a translated field in which only diagonal coefficients
$\sigma^c_{n_cn_c}$ are translated. Indeed noting $d\nu_{diag}(\vec \sigma)$ the diagonal part  of $d\nu(\vec \sigma)$ we have
\bee d\nu_{diag}(\vec \sigma)e^{ i \lambda \sum_c  \sum_{n_c}A(n_c) \sigma^c_{n_c n_c}  + \cD } = \prod_c\prod_{n_c} e^{- \frac{1}{2}  ( \sigma ^c_{n_c n_c}  - i \lambda A(n_c))^2 }.
\ee

Let us define the diagonal operator $D (n, \bar n) =\delta_{n \bar n}  D( n) $ which acts on $\cH$ with eigenvalues 
\bee  D( n) \equiv  \sum_c C(n)  A(n_c) .
\ee
This operator commutes with $C$ since they are both diagonal in the momentum basis. It is bounded 
uniformly in $N $ since from 
\eqref{propdef} and \eqref{Abound} we have 
\bee  \Vert  D( n) \Vert  \le   O(1).\label{Inbound}
\ee
In fact $D$ is also compact as an infinite dimensional operator on $\ell^2 ({\mathbb Z}^3)$, hence  at $N = \infty$, and its square is trace class, since
\bee \sum_{n \in {\mathbb Z}^3} \sum_{c,c'}  \frac{\log (1+ \vert n_c \vert) \log (1+ \vert n_{c'} \vert)  }{(n^2 +1)^2}   \le O(1) . \label{squaretraceclass}
\ee

\begin{figure}[h]
%[!t]
  \begin{center}
  {\includegraphics[width=0.2\textwidth]{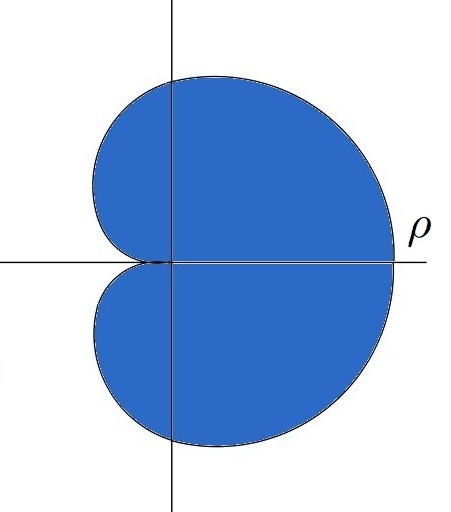}}
   \end{center}
  \caption{A Cardioid Domain}
  \label{cardio}
\end{figure}
 
\begin{lemma} \label{lemmaresbounded} 
For $g$ in the small open cardioid domain $\cC ard_\rho$
defined by $\vert g \vert < \rho \cos [({\rm Arg} \; g )/2]$ (see Figure \ref{cardio}),
the translated resolvent
\bee
R = [\mathbb{I}  -i\lambda C^{1/2}  \vec \sigma C^{1/2} + \lambda^2 D ]^{-1} 
\ee
is well defined and uniformly bounded:
\bee\label{rescardbou}
\Vert R \Vert \le  2  \cos^{-1}( {\rm Arg}\, g /2) .
\ee
\end{lemma}
\prf  In the cardioid domain  we have $\vert {\rm Arg}\, g \vert < \pi$ and for any self-adjoint operator $L$ 
we have 
\bee \Vert ( \mathbb{I}  -i\sqrt g L )^{-1} \Vert \le \cos^{-1}( {\rm Arg}\, g /2).
\ee
Taking  $\rho $ small enough so that $\rho\Vert  D( n) \Vert  < 1/2$, 
the Lemma follows from the power series expansion 
\bee
\Vert ( \mathbb{I}  -i\sqrt g L  +\lambda^2 D )^{-1} \Vert \le \Vert J^{-1} \Vert  \sum_{q=0}^{\infty} \Vert \lambda^2 D J^{-1} \Vert^q \le 2   \cos^{-1}( {\rm Arg}\, g /2)  ,
\ee
with $J=  \mathbb{I}  -i \sqrt g L $.
\qed

\begin{lemma} \label{lemmatrans} For $g$ in the cardioid domain $\cC ard_\rho$, the successive contour translations from $\sigma ^c_{n_c n_c}  $ to
$ \sigma ^c_{n_c n_c}  - i \lambda A(n_c) $ do not cross any singularity of ${\rm \bf Tr} \log_2 \left[\mathbb{I}  -i\lambda C^{1/2}  \vec \sigma C^{1/2}  \right] $.
\end{lemma}
\prf 
To prove that $ {\rm \bf Tr} \log_2 [\mathbb{I}  -i\lambda C^{1/2}  \vec \sigma C^{1/2}]$ is  analytic 
in the combined  translation band of imaginary width $\lambda A(n_c) $ 
for the $\sigma ^c_{n_c n_c}$ variables, one can write
\bee  \log_2 (1-x) =   - \int_0^1 \frac{tx^2}{1-tx} dt
\ee
and then use the previous lemma to prove that, for $g$ in the small open cardioid domain $\cC ard_\rho$, 
the resolvent $R(t)= [\mathbb{I}  -it\lambda C^{1/2}  \vec \sigma C^{1/2} + t\lambda^2 D ]^{-1}$, is also well-defined for any $t \in [0,1]$ by a power series 
of analytic terms uniformly convergent  in the band of of imaginary width $\lambda A(n_c) $. Hence it is analytic in that band. \qed

Hence by Lemma \ref{lemmatrans}
\bee  \label{niceequat3}
Z(g) = \int d\nu(\vec \sigma) e^{- {\rm \bf Tr} \log_2 \left[\mathbb{I}   -i\lambda C^{1/2}  \vec \sigma C^{1/2} + \lambda^2 D \right]  + \cE  } = \int d\nu(\vec \sigma) e^{- V(\sigma)  }  .
\ee
where the $\sigma$ interaction is now defined as
\bee
V(\sigma)  \equiv   {\rm \bf Tr} \log_2 \left[\mathbb{I} -  U \right] - \cE ,  \quad U \equiv i\lambda C^{1/2}  \vec \sigma C^{1/2}   - \lambda^2 D .
\ee

\subsection{Slices and Intermediate Field Representation}
\label{slicesand}

The ``cubic" cutoff $[-N, N]^3$ of the previous section is not very well adapted to the rotation invariant $n^2$ term in the propagator,
nor very convenient for multi-slice analysis as in \cite{MLVE}. In this section we introduce better cutoffs, which are still sharp\footnote{We could also use parametric cutoffs
as in \cite{Riv,BenGeloun:2011rc}, but sharp cutoffs are simpler.}
in the ``momentum space" $\ell_2 ({\mathbb Z})^3$, but not longer factorize over colors. 

It means we fix an integer $M>1$ as ratio of a geometric progression $M^j$
and define the ultraviolet cutoff as a maximal slice index $j_{max}$ so that the previous $N$ roughly corresponds to
$M^{j_{max}}$. 
More precisely, our notation convention is that $\bbone_{x}$ is the characteristic function of the event $x$,
and we define  the following functions of $n \in { \mathbb Z }^3$ : 
\bea
\bbone_{\le 1} &=& \bbone_{1} = \bbone_{1+ n_1^2+n_2^2+n_3^2  \le M^{2} } \\
\bbone_{\le j}&=&  \bbone_{1+ n_1^2+n_2^2+n_3^2  \le M^{2j} } \;\; {\rm for}\; j\ge 2, \\
\bbone_{ j}&=& \bbone_{\le j}  - \bbone_{\le j-1} \;\; {\rm for}\; j\ge 2. 
\label{propmombound}
\eea
(Beware we choose the convention of \emph{lower} indices for slices, as in \cite{MLVE}, not upper
indices as in \cite{Riv}.)

We start with the formulation of the action \eqref{niceequat3} which we have reached in the previous section, and organize it according to the new cutoffs,
so that the previous limit $N \to \infty$ becomes a limit $j_{max} \to \infty$.
The interaction with cutoff $j$ is (since $\bbone_{ j}^2 = \bbone_{ j}$)
\bea
V_{\le j} &\equiv&   {\rm \bf Tr} \log_2 \left[  \mathbb{I} - U_{\le j}  \right]  -  \cE_{\le j} , \\
U_{\le j} &\equiv&   i \lambda \bbone_{\le j}  C^{1/2} {\vec \sigma } C^{1/2} \bbone_{\le j}  - \lambda^2 \bbone_{\le j} D  , \\
\cE_{\le j}&\equiv&   \frac{\lambda^2}{2} \int d\nu(\vec \sigma)  {\rm \bf Tr}  ( \bbone_{\le j} C{\vec \sigma })^2 .
\eea

To define the specific part of the interaction which should be attributed to the scale $j$ we introduce
\bee 
\bbone_{\le j}(t_j) = \bbone_{\le j-1}   + t_j \bbone_{j}
\ee  where $t_j \in [0,1]$ is an interpolation  parameter for the $j$-th scale. Remark that 
\bee \label{theereisasquare}
\bbone^2_{\le j}(t_j) =  \bbone_{\le j-1}   + t^2_j \bbone_{j}.
\ee
The interpolated interaction and resolvents are defined as
\beann
V_{\le j} (t_j) &\equiv &     {\rm \bf Tr} \log_2 \left[  \mathbb{I} - U_{\le j} (t_j) \right] -  \cE_{\le j}  (t_j) 
\ , \\
U_{\le j}(t_j)  &\equiv&   i \lambda \bbone_{\le j} (t_j)  C^{1/2} {\vec \sigma } C^{1/2} \bbone_{\le j}(t_j) -  \lambda^2 \bbone_{\le j} (t_j)  D , \\
\cE_{\le j}   (t_j)  &\equiv&  \frac{\lambda^2}{2} \int d\nu(\vec \sigma)  {\rm \bf Tr}  ( \bbone^2_{\le j}   (t_j) C{\vec \sigma })^2   .
\eeann 

Remark that 
\bee \label{boundcoun}
0 \le   \cE_{\le j}  \le O(1) j  , \quad  0 \le \cE_{\le j}  -   \cE_{\le j-1} \le O(1)  .
\ee

We also define the interpolated resolvent
\bee
R_{\le j}(t_j)  \equiv \frac{1}{  \mathbb{I} - U_{\le j} (t_j) }  .
\ee
When the context is clear, we write simply $V_{\le j} $ for $V_{\le j} (t_j) $, $U_{\le j} $ for $U_{\le j} (t_j) $, $U'$ for $\frac{d}{dt_j}  U_{\le j}  $ and so on. 
We also write $C^{1/2}_{\le j}$ for $  \bbone_{\le j}(t_j) C^{1/2} $,  $C^{1/2}_{j}$ for $  \bbone_{j} C^{1/2} $,
$C_{j}$ for $\bbone_{j} C $, $D_{\le j}$ for $ \bbone_{\le j}(t_j) D$ and $D_{j}$ for $ \bbone_{ j} D $. However beware that we shall write
$C_{\le j}$  for $ \bbone^2_{\le j}(t_j) C$, as this is the natural expression that will always occur in that case. With these notations  we do have the natural relations
\bee  [ÊC^{1/2}_{\le j}  ]^2  = C_{\le j}, \quad  [ C^{1/2}_{j} ]^2 = C_j  .
\ee

We have
\bea  V_{j} &=&  V_{\le j}  - V_{\le j-1} = \int_0^1  dt_j  V'_{\le j} =  \int_0^1  dt_j [ {\rm \bf Tr} \; U'_{\le j}  (  \mathbb{I}  -  R_{\le j} )  - \cE_{\le j}' ]    \nonumber \\
U'_{\le j} &=& i\lambda    C^{1/2}_{j} \vec \sigma C^{1/2}_{\le j} +  i\lambda  C^{1/2}_{\le j}\vec \sigma C^{1/2}_{j}  - \lambda^2 D_j 
\eea
Now we use that $(  \mathbb{I}  -  R_{\le j} ) = - U_{\le j} R_{\le j} = - R_{\le j}U_{\le j}$ and the cyclicity of the trace, plus relations such as $ C^{1/2}_{\le j} C^{1/2}_{j} = t_j  C_j $  to write
\bea {\rm \bf Tr} \; U'_{\le j}  (  \mathbb{I}   -  R_{\le j} )  &=&   - i\lambda  {\rm \bf Tr}  [ÊR_{\le j} U_{\le j}  C^{1/2}_{j} \vec \sigma C^{1/2}_{\le j}  Ê+ R_{\le j} C^{1/2}_{\le j}\vec \sigma C^{1/2}_{j} U_{\le j}    +i  \lambda R_{\le j}U_{\le j} D_j ]  \nonumber \\
&=&  \lambda^2 {\rm \bf Tr}  R_{\le j} [ 2 t_j C^{1/2}_{\le j}\vec \sigma   C_{j} \vec \sigma C^{1/2}_{\le j}  \nonumber  \\
&&   +i \lambda ( D_{\le j} C^{1/2}_{j} \vec \sigma C^{1/2}_{\le j}  + C^{1/2}_{\le j}\vec \sigma C^{1/2}_{j} D_{\le j} ) + U_{\le j} D_j   ]  \label{keyequati}
\\
{\rm \bf Tr} \; \cE'_{\le j}  &=& 2\lambda^2 t_j\int d\nu(\vec \sigma)  {\rm \bf Tr}  ( C_{ j}  {\vec \sigma } C_{\le j}  {\vec \sigma })  .\label{interaj4b}
\eea

Remark that if we replace $R_{\le j}$ by $\mathbb{I} $ in the first term in \eqref{keyequati}  it would exactly cancel the $\cE'_{\le j} $ term. This is nothing but again the exact cancellation
of the last vacuum graph in Figure \ref{divergences} with its counter term.

We now have 
\bee
Z(g, j_{max})= \int d\nu(\vec \sigma) \prod_{j =0}^{j_{\max}}     e^{ - V_j }  . \label{factoredintera}
\ee
As in \cite{MLVE} we define
\bee
W_j(\vec \sigma) = e^{-V_j} -1
\ee
and encode the factorization of the interaction in \eqref{factoredintera} through Grassmann numbers as
\bea\label{eq:partitionfunctionintfield}
Z(g, j_{max}) &=&  \int d\nu  (\vec \sigma) \; \prod_{j =0}^{j_{\max}}     e^{ - V_j } 
=  \int d\nu  (\vec \sigma) \; \Bigl( \prod_{j = 0}^{j_{\max} } d\mu (\bar \chi_j , \chi_j) \Bigr) \; 
 e^{ - \sum_{j = 0}^{j_{\max}}   \bar \chi_j  W_j   ( \vec \sigma)   \chi_j } ,
\eea
where $d \mu(\bar \chi ,\chi ) = d\bar \chi d\chi \; e^{-\bar \chi \chi}$ is the standard normalized Grassmann Gaussian measure with covariance 1.

\section{The Multiscale Loop Vertex Expansion}

We perform now the two-level jungle expansion of \cite{MLVE}. For completeness we summarize the main steps, referring to 
\cite{MLVE} for details.

Considering the set of scales $\cS = [0, j_{max}]$, we denote $\mathbb{I}_{\cS}$ the $\vert \cS \vert $ by $\vert \cS \vert$ identity matrix.
Then we rewrite the partition function as:
\bee
Z(g, j_{max})  =  \int d\nu_\cS \; e^{- W} \; ,
\quad d\nu_{\cS}  = d\nu  ( \sigma  ) \;  d\mu_{\mathbb{I}_\cS} (\{\bar \chi_j , \chi_j\})   \; ,
 \quad W = \sum_{j =0}^{j_{\max}}   \bar \chi_j  W_j   (\vec  \sigma)   \chi_j \; .
\ee 
The first step expands to infinity the exponential of the interaction:
\bee
Z(g, j_{max})   =  \sum_{n=0}^\infty \frac{1}{n!}\int d\nu_{\cS}  \; (-W)^n \; .
\ee
The second step introduces Bosonic replicas for all the vertices in $V = \{1, \cdots , n\}$: 
\bee
Z(g, j_{max}) = \sum_{n=0}^\infty \frac{1}{n!}  \int d\nu_{\cS,V} \;  \prod_{a=1}^n  (-W_a) \; ,
\ee
so that each vertex $W_a$ has now its own set of three Bosonic matrix fields $\vec \sigma^a = \{(\sigma^1)^a, (\sigma^2)^a, (\sigma^3)^a \} $. 
The replicated measure is completely degenerate between replicas (each of the three colors remaining independent of the others):
\bee
d\nu_{\cS,V}  = d\nu_{ \bbone_V} (\{  \vec \sigma^a\}) \;  d\mu_{\mathbb{I}_\cS  } (\{\bar \chi_j , \chi_j\})   \; ,\quad 
W_a =  \sum_{j =0}^{j_{\max}}   \bar \chi_j W_j   (  \vec \sigma^a )  \chi_j \; .
\ee

The obstacle to factorize the functional integral $Z$ over vertices and to compute $\log Z$ lies in the Bosonic degenerate blocks $\bbone_V$ and 
in the Fermionic fields. In order to remove this obstacle we need to apply {\it two} successive
forest formulas \cite{BK,AR1}, one Bosonic, the other Fermionic. The main difference with \cite{MLVE} is that the Bosonic
forest will be \emph{three-colored} since there are three colors for the intermediate matrix fields.

To analyze the block $\bbone_V$ in the measure $d \nu$ 
we introduce coupling parameters $x_{ab}=x_{ba}, x_{aa}=1$ between the Bosonic vertex replicas. Since there are three colors, and
since the interpolation
parameters are color-blind,
we obtain a sum over three-colored forests. Representing Gaussian integrals as derivative operators
as in \cite{MLVE} we have
\bee
Z(g, j_{max}) = \sum_{n=0}^\infty \frac{1}{n!} 
\Bigl[ e^{\frac{1}{2} \sum_{a,b=1}^n x_{ab} 
 \sum_{c =1}^3 \frac{\partial}{\partial (\sigma^c)^a}\frac{\partial}{\partial (\sigma^c)^b} 
   +  \sum_{j=0}^{j_{\max}} \frac{\partial}{\partial \bar \chi_j  } \frac{\partial}{\partial \chi_j } } \; 
   \prod_{a=1}^n \Bigl( -  \sum_{j =0}^{j_{\max}}   \bar \chi_j W_j   (  \vec \sigma^a )  \chi_j \Bigr) 
   \Bigr]_{ \genfrac{}{}{0pt}{}{ \vec \sigma, \chi,  \bar\chi  =0}{x_{ab}=1 }} \;.
\ee
The third step applies the standard Taylor forest formula of \cite{BK,AR1} to the $x$ parameters. 
We denote by $\cF^{3c}_{B}$ a three-colored Bosonic forest with $n$ vertices labelled $\{1,\dots n\}$.
It means an acyclic set of edges over $V$ in which each edge  $\ell_{B}$ has a specific color $c(\ell) \in \{1,2,3\}$. For
$\ell_{B}$ a generic edge of the forest we denote by $a(\ell_B), b(\ell_B)$ the end vertices of $\ell_B$. The result of the Taylor forest formula is:
\bea
&&  Z(g, j_{max})  = \sum_{n=0}^\infty \frac{1}{n!} \sum_{\cF^{3c}_{B}} \int_{0}^1 \Bigl( \prod_{\ell_B\in \cF^{3c}_B } dw_{\ell_B} \Bigr)
\; \;  \Bigg[  e^{\frac{1}{2} \sum_{a,b=1}^n X_{ab}(w_{\ell_B})  
   \sum_{c =1}^3 \frac{\partial}{\partial (\sigma^c)^a}\frac{\partial}{\partial (\sigma^c)^b} 
   +  \sum_{j=0}^{j_{\max}} \frac{\partial}{\partial \bar \chi_j  } \frac{\partial}{\partial \chi_j  } }
\crcr
&& \qquad \qquad \qquad \times 
   \prod_{\ell_B \in \cF^{3c}_{B}} \Bigl(  
   \frac{\partial}{\partial (\sigma^{c( \ell_B) })^{a(\ell_B)}}\frac{\partial}{\partial (\sigma^{c(\ell_B)})^{b(\ell_B)}}  \Bigr) \;
 \prod_{a=1}^n \Bigl( - \sum_{j =0}^{j_{\max}}   \bar \chi_j W_j   (  \vec \sigma^a )  \chi_j \Bigr) \Bigg]_{\vec \sigma ,\chi , \bar\chi =0 } \; ,
\eea
where $X_{ab}(w_{\ell_B})$ is the infimum over the parameters $w_{\ell_B}$ in the unique path
in the forest $\cF^{3c}_B$ connecting $a$ to $b$. This infimum is set to 
$1$ if $a=b$ and to zero if $a$ and $b$ are not connected by the forest \cite{BK,AR1}.

The colored forest $\cF^{3c}_B$ partitions the set of vertices into blocks $\cB$ corresponding to its connected components. In each such block the 
edges of $\cF^{3c}_B$ form a spanning tree. Remark that such blocks can be reduced to single vertices. Any vertex $a$
belongs to a unique Bosonic block $\cB$.
Contracting every Bosonic block to an ``effective vertex'' we obtain a reduced set which we denote $\{n\}/\cF_B$.

\begin{figure}[!t]
  \begin{center}
  {\includegraphics[width=0.5\textwidth]{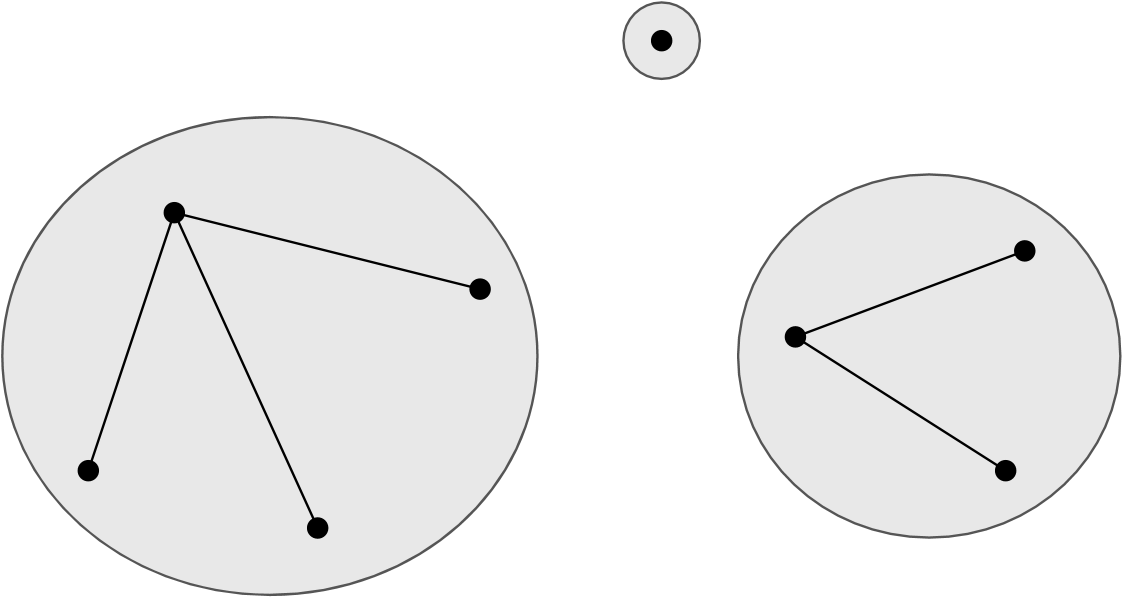}}
   \end{center}
  \caption{A Bosonic forest, the bosonic blocks $\cB$ are represented in gray.}
  \label{forest1}
\end{figure}

\begin{figure}[!t]
  \begin{center}
  {\includegraphics[width=0.5\textwidth]{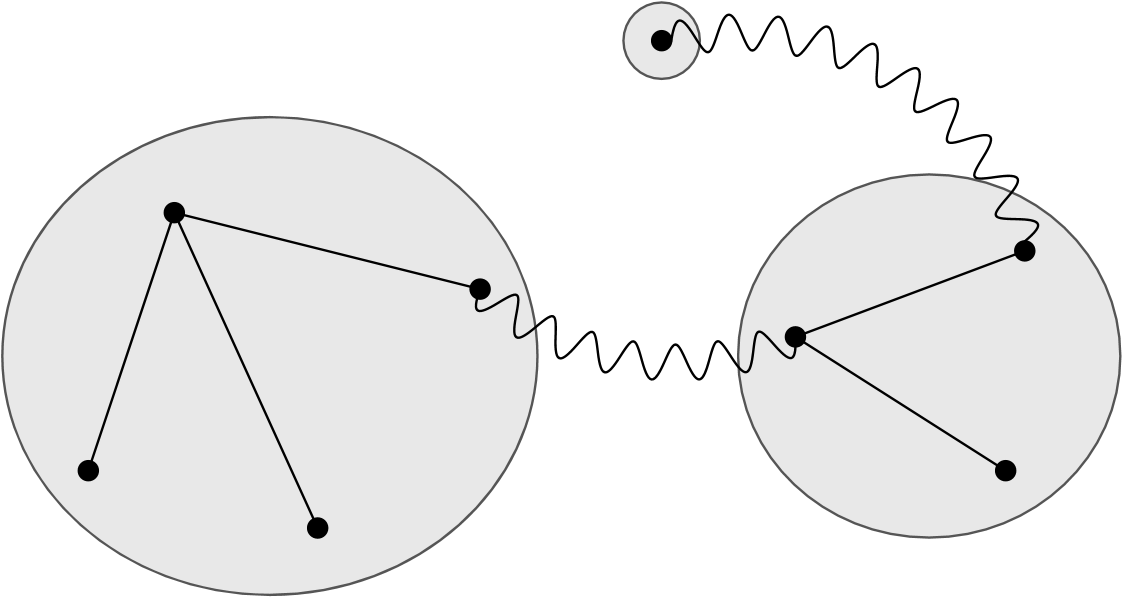}}
   \end{center}
  \caption{A two level jungle, the fermionic (wiggly) edges can only link two different bosonic blocks. }
  \label{jungle1}
\end{figure}

The fourth step introduces replica Fermionic fields $\chi^{\cB}_j$ for these blocks of $\cF^{3c}_B$ (i.e. for the effective vertices 
of $\{n\}/\cF^{3c}_B$) and replica coupling parameters $y_{\cB\cB'}=y_{\cB'\cB}$. 
The fifth and last step applies (once again) the forest formula, 
this time for the $y$'s, leading to a set of Fermionic edges $\cL_F$ forming an (uncolored) forest 
in $\{n\}/\cF^{3c}_B$ (hence connecting Bosonic blocks). Denoting $L_{F} $ a generic Fermionic edge connecting blocks and $\cB(L_F), \cB'(L_F) $ 
the end blocks of the Fermionic edge $L_F$ we
follow exactly the same steps than in \cite{MLVE} 
and obtain a two level-jungle formula \cite{AR1}Ê
in which the first level is three-colored and the second level is uncolored.
The result writes
\bee
Z(g, j_{max})  =  \sum_{n=0}^\infty \frac{1}{n!}  \sum_{\cJ} \;\sum_{j_1=0}^{j_{\max }  } 
  \dots \sum_{j_n=0}^{j_{\max} }
 \;  \int dw_\cJ  \;  \int d\nu_{ \cJ}  
\quad   \partial_\cJ   \Big[ \prod_{\cB} \prod_{a\in \cB}   \Bigl(    W_{j_a}   (  \vec \sigma^a )  
\chi^{ \cB }_{j_a}   \bar \chi^{\cB}_{j_a}  \Bigr)  \Big] \; ,
\ee
where
\begin{itemize}

\item the sum over $\cJ$ runs over all two-level jungles, the first level of which is three-colored,
hence over all ordered pairs $\cJ = (\cF^{3c}_B, \cF_F)$ of two (each possibly empty) 
disjoint forests on $V$, such that $\cF^{3c}_B$ is a three colored-forest, $\cF_F$ is an uncolored forest and
$\bar \cJ = \cF^{3c}_B \cup \cF_F $ is still a forest on $V$. The forests $\cF_B^{3c}$ and $\cF_F$ are the Bosonic and Fermionic components of $\cJ$.
Bosonic edges $\ell_B \in \cF^{3c}_B$ have a well-defined color $c(\ell) \in \{1,2,3\}$ 
and Fermionic edges $\ell_F \in \cF_F$ are uncolored.
 
\item  $\int dw_\cJ$ means integration from 0 to 1 over parameters $w_\ell$, one for each edge $\ell \in \bar\cJ$, namely
$\int dw_\cJ  = \prod_{\ell\in \bar \cJ}  \int_0^1 dw_\ell  $.
There is no integration for the empty forest since by convention an empty product is 1. A generic integration point $w_\cJ$
is therefore made of $\vert \bar \cJ \vert$ parameters $w_\ell \in [0,1]$, one for each $\ell \in \bar \cJ$.

\item 
\bee \partial_\cJ  = \prod_{\genfrac{}{}{0pt}{}{\ell_B \in \cF^{3c}_B}{\ell_B=(a,b)}} \Bigl(
\frac{\partial}{\partial (\sigma^{c(\ell_B)})^a}\frac{\partial}{\partial (\sigma^{c(\ell_B)})^b} \Bigr)
\prod_{\genfrac{}{}{0pt}{}{\ell_F \in \cF_F}{\ell_F=(d,e) } } \delta_{j_{d } j_{e } } \Big(
   \frac{\partial}{\partial \bar \chi^{\cB(d)}_{j_{d}  } }\frac{\partial}{\partial \chi^{\cB(e)}_{j_{e}  } }+ 
    \frac{\partial}{\partial \bar \chi^{ \cB( e) }_{j_{e} } } \frac{\partial}{\partial \chi^{\cB(d) }_{j_{d}  } }
   \Big) \; ,
\ee
where $ \cB(d)$ denotes the Bosonic block to which the vertex $d$ belongs. 

\item The measure $d\nu_{\cJ}$ has covariance $ X (w_{\ell_B}) \otimes \bbone_\cS $ on Bosonic variables and $ Y (w_{\ell_F}) \otimes \mathbb{I}_\cS  $  
on Fermionic variables, hence
\bee
\int d\nu_{\cJ} F = \biggl[e^{\frac{1}{2} \sum_{a,b=1}^n X_{ab}(w_{\ell_B })  \sum_{c =1}^3 \frac{\partial}{\partial (\sigma^c)^a}\frac{\partial}{\partial (\sigma^c)^b} 
   +  \sum_{\cB,\cB'} Y_{\cB\cB'}(w_{\ell_F})\sum_{a\in \cB,  b\in \cB' } \delta_{j_aj_b}
   \frac{\partial}{\partial \bar \chi_{j_a}^{\cB} } \frac{\partial}{\partial \chi_{j_b}^{\cB'} } }   F \biggr]_{\sigma = \bar\chi =\chi =0}\; .
\ee

\item  $X_{ab} (w_{\ell_B} )$  is the infimum of the $w_{\ell_B}$ parameters for all the Bosonic edges $\ell_B$
in the unique path $P^{\cF_B}_{a \to b}$ from $a$ to $b$ in $\cF_B$. The infimum is set to zero if such a path does not exists and 
to $1$ if $a=b$. 

\item  $Y_{\cB\cB'}(w_{\ell_F})$  is the infimum of the $w_{\ell_F}$ parameters for all the Fermionic
edges $\ell_F$ in any of the paths $P^{\cF_B \cup \cF_F}_{a\to b}$ from some vertex $a\in \cB$ to some vertex $b\in \cB'$. 
The infimum is set to $0$ if there are no such paths, and to $1$ if such paths exist but do not contain any Fermionic edges.

\end{itemize}

Remember that a main property of the forest formula is that the symmetric $n$ by $n$ matrix $X_{ab}(w_{\ell_B})$ 
is positive for any value of $w_\cJ$, hence the Gaussian measure $d\nu_{\cJ} $ is well-defined. The matrix $Y_{\cB\cB'}(w_{\ell_F})$
is also positive, with all elements between 0 and 1. Since the slice assignments, the fields, the measure and the integrand are now 
factorized over the connected components of $\bar \cJ$, the logarithm of $Z$ is easily computed as exactly the same sum but restricted 
to two-levels spanning trees (whose first level is three-colored):
\bea \label{treerep}  
\log Z(g, j_{max})=  \sum_{n=1}^\infty \frac{1}{n!}  \sum_{\cJ \;{\rm tree}} \;\sum_{j_1=1}^{j_{\text{max}}} 
  \dots \sum_{j_n=1}^{j_{\text{max}} }
 \;  \int dw_\cJ  \;  \int d\nu_{ \cJ}  
\quad   \partial_\cJ   \Big[ \prod_{\cB} \prod_{a\in \cB}   \Bigl(    W_{j_a}   (  \vec \sigma^a )  
 \chi^{ \cB }_{j_a} \bar \chi^{\cB}_{j_a} \Bigr)    \Big] \; , 
\eea
where the sum is the same but conditioned on $\bar \cJ = \cF^{3c}_B \cup \cF_F$ being a \emph{spanning tree} on $V= [1, \cdots , n]$.

Our main result is
\begin{theorem} \label{thetheorem} Fix  $\rho >0$ small enough.
The series \eqref{treerep} is absolutely and uniformly in $j_{max}$ convergent for $g$ in the small open cardioid domain $\cC ard_\rho$
defined by $\vert g \vert < \rho \cos [({\rm Arg} \; g )/2]$ (see Figure \ref{cardio}).
Its ultraviolet limit $\log Z (g) = \lim_{j_{max}  \to \infty}  \log Z(g, j_{max})$ is therefore well-defined and
analytic in that cardioid domain; furthermore it is the Borel sum
of its perturbative series in powers of $g$. 
\end{theorem}

The rest of the paper is devoted to the proof of this Theorem.

\section{The Bounds}

\subsection{Grassmann Integrals}

The Grassmann Gaussian part of the functional integral \eqref{treerep} is also treated exactly as in \cite{MLVE}, resulting in the same computation:
\bea\label{eq:grassmaint}
\int \prod_{\cB} \prod_{a\in \cB}  ( d  \bar \chi^{\cB}_{j_a}  d  \chi^{\cB}_{j_a}   ) 
 e^{  - \sum_{a,b=1}^n    \bar \chi^{\cB(a)}_{j_a} {\bf Y}_{ab}  \chi^{\cB(b)}_{j_b}  }
\prod_{\genfrac{}{}{0pt}{}{\ell_F \in \cF_F}{\ell_F=(a,b) } } \delta_{j_{a } j_{b } } \Big(
    \chi^{\cB(a)}_{j_{a}  } \bar  \chi^{\cB(b)}_{j_{b } }  + 
     \chi^{ \cB( b) }_{j_{b} }  \bar  \chi^{\cB(a) }_{j_{a} } \Big) \nonumber
 \\ =   
 \Bigl( \prod_{\cB} \prod_{\genfrac{}{}{0pt}{}{a,b\in \cB}{a\neq b}} (1-\delta_{j_aj_b}) \Bigr)
 \Bigl( \prod_{\genfrac{}{}{0pt}{}{\ell_F \in \cF_F}{\ell_F=(a,b) } } \delta_{j_{a } j_{b } } \Bigr)
 \Bigl( {\bf Y }^{\hat b_1 \dots \hat b_k}_{\hat a_1 \dots \hat a_k}  + 
 {\bf Y }^{\hat a_1 \dots \hat b_k}_{\hat b_1 \dots \hat a_k}+\dots + {\bf Y }_{\hat b_1 \dots \hat b_k}^{\hat a_1 \dots \hat a_k}   \Bigr) \; ,
\eea 
where $k= \vert  \cF_F \vert $, the sum runs over the $2^k$ ways to exchange an $a_i$ and a $b_i$,
and the $Y$ factors are (up to a sign) the minors of $Y$ with the lines $b_1\dots b_k$ and the columns $a_1\dots a_k$ deleted.
The most important factor in \eqref{eq:grassmaint} is 
$\Bigl( \prod_{\cB} \prod_{\genfrac{}{}{0pt}{}{a,b\in \cB}{a\neq b}} (1-\delta_{j_aj_b}) \Bigr)$
which means that the scales obey to a \emph{hard core constraint inside each block}.
Positivity of the $Y$ covariance means as usual that the $Y$ minors are all bounded by 1 \cite{AR2,MLVE}, namely
for any $a_1,\dots a_k$ and $b_1,\dots b_k$,
 \bee
   \Big{|}  {\bf Y }^{\hat a_1 \dots \hat b_k}_{\hat b_1 \dots \hat a_k} \Big{|}\le 1 \; .
 \ee

\subsection{Bosonic Integrals}

The main problem is now the evaluation of the Bosonic integral in \eqref{treerep}. Since it factorizes over the Bosonic blocks,
it is sufficient to bound separately this integral in each fixed block $\cB$.
In such a block the Bosonic forest $\cF^{3c}_B$ restricts to a three-colored Bosonic tree 
$\cT^{3c}_{\cB}$, 
and the Bosonic Gaussian measure  $d \nu $ restricts to $d \nu_\cB$ defined by
\bee
\int d \nu_\cB  F_\cB = \biggl[ e^{\frac{1}{2} \sum_{a,b\in \cB} X_{ab}(w_{\ell_B }) 
\sum_{c =1}^3 \frac{\partial}{\partial (\sigma^c)^a}\frac{\partial}{\partial (\sigma^c)^b}} F_\cB \biggr]_{\sigma =0}.
\ee
The Bosonic integrand $F_\cB =  \prod_{\ell \in \cT^{3c}_{\cB} , \ell_B=(a,b)} \Bigl(
\frac{\partial}{\partial (\sigma^{c(\ell_B)})^a}\frac{\partial}{\partial (\sigma^{c(\ell_B)})^b} \Bigr) \prod_{a\in \cB}   \Bigl(    W_{j_a}   (  \vec \sigma^a ) \Bigr)
$ can be written in shorter notations as 
\bee   F_\cB =  \prod_{a\in \cB}    \bigl[ \prod_{s \in S^a_\cB} \partial \sigma_s W_{j_a}   \bigr]
\ee
where $S^a_\cB$ runs over the set of all edges in $\cT^{3c}_{\cB}$ which end at vertex $a$, hence $\vert S^a_\cB \vert = d_a( \cT^{3c}_{\cB} ) $, the
degree or coordination of the tree $\cT^{3c}_{\cB}$ at vertex $a$. To each element $s$ is therefore associated a well-defined color and 
well-defined matrix elements
(which have to be summed later after identifications are made through the edges of $\cT^{3c}_{\cB}$).

When $\cB$ has more than one vertex,
since $\cT^{3c}_{\cB}$ is a tree, each vertex $a \in \cB$ is touched by at least one derivative and we can replace 
$W_{j_a} =e^{- V_{j_a}} -1$ by $ e^{- V_{j_a}}$ (the derivative of 1 giving 0) and write
\bee \label{manyder}  F_\cB =    \prod_{a\in \cB}    \bigl[ \prod_{s \in S^a_\cB} \partial \sigma_s  e^{- V_{j_a}}  \bigr].
\ee
We can evaluate the derivatives in \eqref{manyder} through the Fa\`a di Bruno formula:
\bee
\prod_{s\in S} \partial \sigma_s   f\bigl( g( \sigma ) \bigr) =   \sum_{\pi } f^{|\pi|}\bigl( g( \sigma ) \bigr) \prod_{b\in \pi}  \left(\bigl[ \prod_{s\in b} \partial \sigma_s \bigr] g (\sigma)\right) \; ,
\ee
where $\pi$ runs over the partitions of the set $S$ and $b$ runs through the blocks of the partition $\pi$. In our case $f$, the exponential
function, is its own derivative, hence the formula simplifies to
\bea F_\cB&=&   \prod_{a\in \cB}   e^{- V_{j_a}}   \biggl[  \sum_{\pi^a} \prod_{b^a\in \pi^a} \;    \bigl[\prod_{s\in b^a} \partial \sigma_s\bigr]  (-V_{j_a})  \biggr]\; ,
\label{partitio}
\eea
where $\pi^a$ runs over partitions of $S^a_\cB$ into blocks $b^a$.

We recall that, with the notations of Section \ref{slicesand}
\bea  V_{j} &=& \lambda^2 \int_0^1  dt_j   \biggl(
{\rm \bf Tr} R_{\le j}  \biggl[ 2 t_j C^{1/2}_{\le j}\vec \sigma   C_{j} \vec \sigma C^{1/2}_{\le j}     +i \lambda   ( D_{\le j}C^{1/2}_{j} \vec \sigma C^{1/2}_{\le j}  + C^{1/2}_{\le j}\vec \sigma C^{1/2}_{j} D_{\le j})  + U_{\le j} D_j  \biggr] 
\nonumber\\ && -2 t_j \int d\nu(\vec \sigma)   C_{ j}  {\vec \sigma } C_{\le j}  {\vec \sigma } \biggr) 
\nonumber\\
&=& \cD_j  + 2\lambda^2 \int_0^1 t_j  dt_j \biggl[ 
{\rm \bf Tr} R_{\le j}     C^{1/2}_{\le j}\vec \sigma   C_{j} \vec \sigma C^{1/2}_{\le j}     - \int d\nu(\vec \sigma)   C_{ j}  {\vec \sigma } C_{\le j}  {\vec \sigma }    \biggr]  ,\label{interaj4c}
\eea
where $\cD_j$ gathers all terms with a $D$ factor:
\bea  \cD_j &\equiv&  \lambda^2 \int_0^1  dt_j   {\rm \bf Tr} \, R_{\le j}  \bigl[ i \lambda ( D_{\le j}  C^{1/2}_{j} \vec \sigma C^{1/2}_{\le j}  
+ C^{1/2}_{\le j}\vec \sigma C^{1/2}_{j}  D_{\le j} )  + U_{\le j} D_j ] \nonumber\\
&=& i\lambda^3 \int_0^1  dt_j   {\rm \bf Tr} \, R_{\le j} \bigl[ D_{\le j}  C^{1/2}_{j} \vec \sigma C^{1/2}_{\le j}  +C^{1/2}_{\le j}\vec \sigma C^{1/2}_{j}  D_{\le j}  +  C^{1/2}_{\le j}  \vec \sigma C^{1/2}_{\le j}D_j   +i\lambda D_{\le j}   D_j \bigr] .
\label{defDj}
\eea
The last term in \eqref{interaj4c} is the $\cE'_{\le j} $  constant term, which does not depend on $\vec \sigma$. 
Hence remembering that $\partial \sigma_s$  really stand for a derivative with well defined color and matrix elements $\partial \sigma^{c_s}_{n^{c_s},\bar n^{c_s}}$, we get
following the order of the terms in \eqref{interaj4c}
\bea
  \partial \sigma_1 ( -V_j ) &=&  i\lambda^3\int_0^1  dt_j  \biggl( {\rm \bf Tr} \, i\lambda R_{\le j}C^{1/2}_{\le j}\Delta^1 C^{1/2}_{\le j}R_{\le j} \nonumber\\
&&\qquad \times \bigl[ t_j^2 D_{ j}  C^{1/2}_{j} \vec \sigma C^{1/2}_{\le j}  + t^2_j C^{1/2}_{\le j}\vec \sigma C^{1/2}_{j}  D_{ j}  +  t_j C^{1/2}_{\le j}  \vec \sigma C^{1/2}_{j}D_j   +i\lambda t_j^2 D_{ j}   D_j \bigr] \nonumber\\
&&\qquad +{\rm \bf Tr}R_{\le j} \bigl[ t_j^2 D_{j}  C^{1/2}_{j} \Delta^1  C^{1/2}_{\le j}  +t^2_j C^{1/2}_{\le j}\Delta^1  C^{1/2}_{j}  D_{ j}  + t_j C^{1/2}_{\le j}  \Delta^1  C^{1/2}_{ j}D_j   \bigr]\biggr)\nonumber\\
&+& 2\lambda^2 \int_0^1 t_j  dt_j \ i\lambda {\rm \bf Tr}R_{\le j}C^{1/2}_{\le j}\Delta^1 C^{1/2}_{\le j}R_{\le j}\left[   C^{1/2}_{\le j}\vec \sigma   C_{j} \vec \sigma C^{1/2}_{\le j}\right]\\ \label{developcycle}
&&\qquad + {\rm \bf Tr} R_{\le j}  \left[   C^{1/2}_{\le j}\Delta^1   C_{j} \vec \sigma C^{1/2}_{\le j} +  C^{1/2}_{\le j}\vec \sigma   C_{j} \Delta^1 C^{1/2}_{\le j}  \right] \ .\nonumber\eea
The formula for several successive derivations is similar and straightforward although longer:
\bea
  \prod_{s=1}^k\partial \sigma_s (-V_j ) &=& \sum_{\tau} i\lambda^3\int_0^1  dt_j  \biggl( {\rm \bf Tr} \,\bigl[ \prod_{s=1}^{k} i\lambda R_{\le j} C^{1/2}_{\le j} \Delta^{\tau (s)} C^{1/2}_{\le j}] R_{\le j} \label{developcycles}\\
&&\times \bigl[ t_j^2 D_{ j}  C^{1/2}_{j} \vec \sigma C^{1/2}_{\le j}  +t^2_j C^{1/2}_{\le j}\vec \sigma C^{1/2}_{j}  D_{ j}  + t_j C^{1/2}_{\le j}  \vec \sigma C^{1/2}_{ j}D_j   +i\lambda t_j D_{ j}   D_j \bigr] \nonumber\\
&& +{\rm \bf Tr}\bigl[ \prod_{s=2}^{k} i\lambda R_{\le j} C^{1/2}_{\le j} \Delta^{\tau (s)} C^{1/2}_{\le j}] R_{\le j}\nonumber\\
&&\times \bigl[t_j^2 D_{ j}  C^{1/2}_{j} \Delta^{\tau (1)}  C^{1/2}_{\le j}  + t^2_j C^{1/2}_{\le j}\Delta^{\tau (1)}  C^{1/2}_{j}  D_{ j}  +  t_j C^{1/2}_{\le j}  \Delta^{\tau (1)}  C^{1/2}_{ j}D_j   \bigr]\biggr)\nonumber\\
&+& 2\lambda^2 \int_0^1 t_j  dt_j   {\rm \bf Tr}\bigl[ \prod_{s=1}^{k} i\lambda R_{\le j} C^{1/2}_{\le j} \Delta^{\tau (s)} C^{1/2}_{\le j}]R_{\le j}\left[   C^{1/2}_{\le j}\vec \sigma   C_{j} \vec \sigma C^{1/2}_{\le j}\right]\nonumber\\
&&+{\rm \bf Tr}\bigl[ \prod_{s=2}^{k} i\lambda R_{\le j} C^{1/2}_{\le j} \Delta^{\tau (s)} C^{1/2}_{\le j}] R_{\le j}  \left[   C^{1/2}_{\le j}\Delta^{\tau (1)}   C_{j} \vec \sigma C^{1/2}_{\le j} +  C^{1/2}_{\le j}\vec \sigma   C_{j} \Delta^{\tau (1)} C^{1/2}_{\le j} \right] \nonumber\\
&&+{\rm \bf Tr}\bigl[ \prod_{s=3}^{k} i\lambda R_{\le j} C^{1/2}_{\le j} \Delta^{\tau (s)} C^{1/2}_{\le j}] R_{\le j}  \left[   C^{1/2}_{\le j}\Delta^{\tau (1)}   C_{j} \Delta^{\tau (2)}  C^{1/2}_{\le j} +  C^{1/2}_{\le j} \Delta^{\tau (2)}    C_{j} \Delta^{\tau (1)} C^{1/2}_{\le j} \right]\ . \nonumber
 \eea
 We used $C^{1/2}_{ j}D_{\le j} =t_j C^{1/2}_{\le j}D_{ j} = t_j^2 C^{1/2}_{ j}D_j $. In \eqref{developcycles} the sum over $\tau $ runs over the
 permutations of $[1,k]$ and $\Delta^s$, defined as
 $(\Delta^s)_{m \bar m} = \frac{\partial  \vec \sigma_{m\bar m}  }{\partial \sigma^{c_s}_{n^{c_s},\bar n^{c_s}}} $,
 is the tensor product of the identity matrix on colors $c' \ne c_s$ 
with the matrix  on color $c_s$ with zero entries everywhere except at position $n^{c_s},\bar n^{c_s} $ where it has entry one.
These formulas express the derivatives of the trace as a sum over all cycles with exactly $k$ derivatives, zero or one operator $D$, and 
\emph{up to two remaining numerator $\sigma$ fields} (one at most if the cycle contains an operator $D$). Rewriting $D_j$ as
\bea
D_j = C^{1/2}_{ j} A_j C^{1/2}_{ j}\, ,\qquad (A_j)_{n\bar n} = {\bf 1}_j(n)A(n) \delta_{n\bar n}\ ,
\eea
the cycles have $2k$ to $2k+4$ numerator half-propagators$C^{1/2}$, 
and \emph{two of them must form a $C_j$, hence exactly a propagator of scale $j$}. 

The Bosonic integral in a block can be written therefore in a simplified manner as:
\bea\label{eq:bosogauss}
&& \int d \nu_\cB  F_\cB = \sum_G \int d \nu_\cB \prod_{a\in \cB}  e^{-V_{j_a} (\sigma_a) }  A_G (\sigma)
 \; ,
\eea 
where we gather the result of the derivatives as a sum over graphs $G$ of corresponding amplitudes $A_G( \sigma)$.

These graphs $G$ are still forests, with effective loop vertices\footnote{We recall that loop vertices are the traces obtained by $\sigma$ derivatives acting on the intermediate
field action \cite{Rivasseau:2007fr}.}, one for each $b^a \in \pi^a, a \in \cB$,  
each of them expressed as a trace of a product of three-stranded operators 
by \eqref{developcycles}, with $k = \vert b^a\vert$.
Each such effective vertex of $G$ bears \emph{at most two} $\sigma$ insertions plus 
exactly $\vert b^a \vert$ $\Delta$ insertions, which are contracted together via the colored edges of the tree $\cT^{3c}_\cB$. 

We define the corners of the graph as the pair made of two consecutive insertions of either $\Delta$,  $\sigma$ or  $A_j$ operators.  
Then, each corner of the vertices bears a $C^{1/2}_{(\le) j_a}R_{\le j_a} C^{1/2}_{(\le) j_a}$ operator (the $C^{1/2}$'s being either $C^{1/2}_j$ or $C^{1/2}_{\le j}$), except one distinguished corner which bears a $C_{j_a}$ operator and no resolvent.

Note that to each initial $W_{j_a}$ may correspond several  effective loop vertices $V_{b^a}$,
depending of the partitioning of $S^a_\cB$ in \eqref{partitio}. Therefore although
at fixed $|\cB|$ the number of (colored) edges $E(G)$ for any $G$ in the sum \eqref{eq:bosogauss} is exactly $|\cB|-1$,
the number of connected components $c(G)$ 
is not fixed but simply bounded by $|\cB|-1$ (each edge can belong to a single connected component). Similarly
the number $V(G)= c(G) + E(G)$ of effective loop vertices of $G$ is not fixed, and simply obeys the bounds
\bee   \vert \cB \vert  \le V(G) \le 2(|\cB|-1) . \label{foreboun}
\ee

From now on we shall simply call ``vertices" the effective loop vertices of $G$, as we shall no longer meet the initial $W_{j_a}$ vertices.

\begin{figure}[!h]
\begin{center}
{\includegraphics[width=4.5cm]{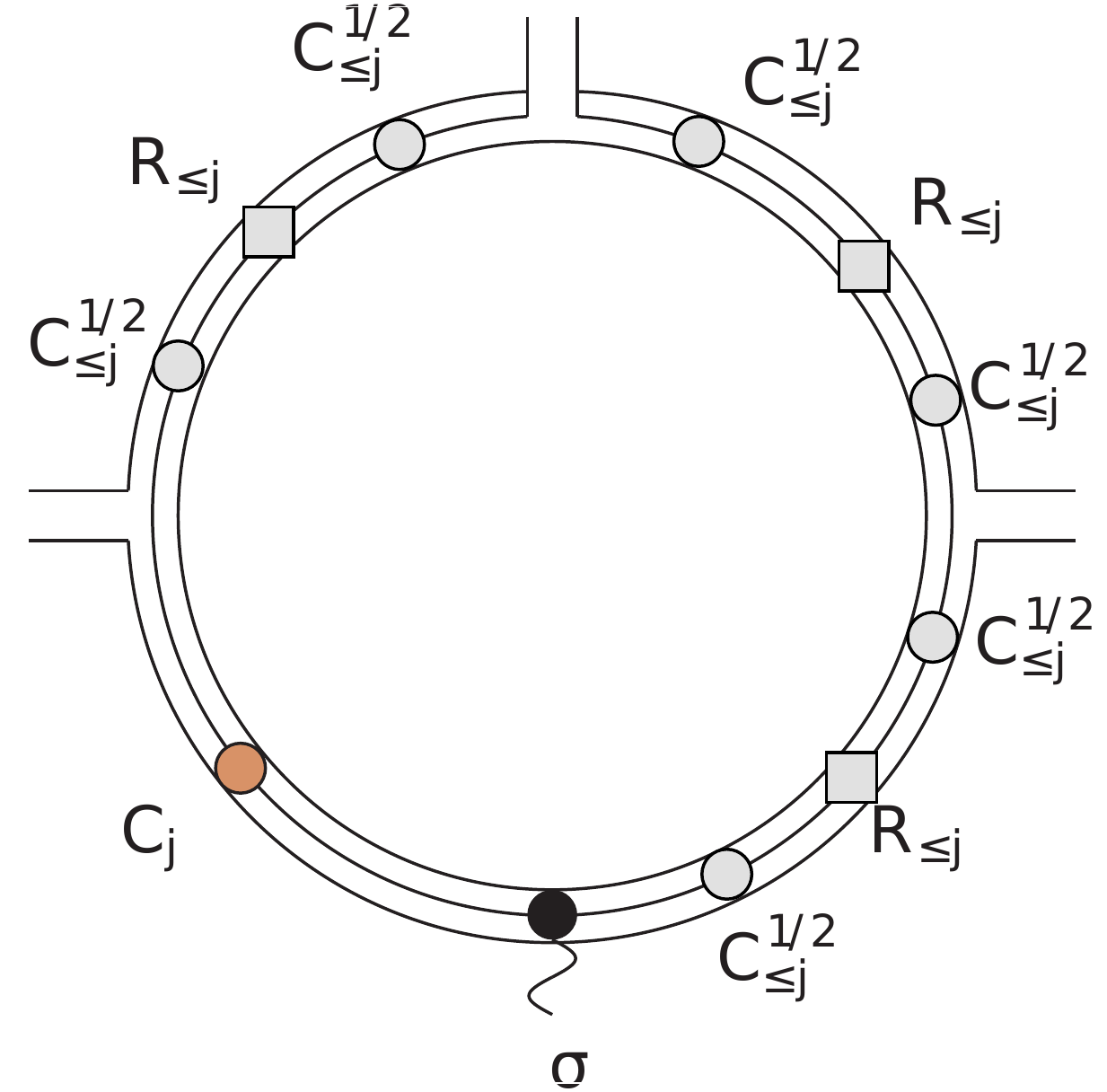}}
\end{center}
\caption{A detailed three-stranded vertex with its cycle of operators. Open strands corresponds to $\Delta$ operators, and to half edges of the tree $\cT^{3c}_\cB$.}
\label{vertex}
\end{figure}

When the block $\cB$ is reduced to a single vertex $a$, we have a simpler contribution for which
an important cancellation occurs due to the presence of the logarithmically divergent counter term in $V_j$.
More precisely (writing simply $j$ for $j_a$)
\bea
\int d \nu_\cB  F_\cB  &=&  \int d \nu (\vec \sigma)  \bigl[ e^{- V_{j}(\vec \sigma) } - 1\bigr] = \int_0^1 dt  
\int d \nu (\vec \sigma)   V_{j}(\vec \sigma)  e^{- tV_{j}(\vec \sigma) }\nonumber\\
&=& \int_0^1 dt  
\int d \nu (\vec \sigma)\cD_je^{- tV_{j}(\vec \sigma) }\label{singbloc} \\
&+&2\lambda^2\int_0^1 dt \int_0^1   dt_j 
\int d \nu (\vec \sigma) t_j  \biggl[ 
{\rm \bf Tr} R_{\le j}     C^{1/2}_{\le j}\vec \sigma   C_{j} \vec \sigma C^{1/2}_{\le j}     - \int d\nu(\vec \sigma)   C_{ j}  {\vec \sigma } C_{\le j}  {\vec \sigma }    \biggr]
e^{- tV_{j}(\vec \sigma) }\ ,
\nonumber\eea
and, using $R_{\le j} = 1 +  i \lambda R_{\le j}  (i \lambda D_{\le j} + C^{1/2}_{\le j}  \vec \sigma C^{1/2}_{\le j} ) $,
\bea
&&\int d \nu (\vec \sigma) t_j  \biggl[ 
{\rm \bf Tr} R_{\le j}     C^{1/2}_{\le j}\vec \sigma   C_{j} \vec \sigma C^{1/2}_{\le j}     - \int d\nu(\vec \sigma)   C_{ j}  {\vec \sigma } C_{\le j}  {\vec \sigma }    \biggr]
e^{- tV_{j}(\vec \sigma) }\nonumber\\
&=& \int d \nu (\vec \sigma) t_j  \biggl[ 
 i\lambda{\rm \bf Tr} R_{\le j}  C^{1/2}_{\le j}  \vec \sigma   C_{\le j}\vec \sigma   C_{j} \vec \sigma C^{1/2}_{\le j} 
-\lambda^2 {\rm \bf Tr} R_{\le j}  D_{\le j}   C^{1/2}_{\le j}\vec \sigma   C_{j} \vec \sigma C^{1/2}_{\le j} \nonumber\\
&&\qquad\qquad +\ 
{\rm \bf Tr}  C^{1/2}_{\le j}\vec \sigma   C_{j} \vec \sigma C^{1/2}_{\le j}     - \int d\nu(\vec \sigma)   C_{ j}  {\vec \sigma } C_{\le j}  {\vec \sigma }    \biggr]
e^{- tV_{j}(\vec \sigma) }\nonumber\\
&=& \int d \nu (\vec \sigma) t_j  \biggl[ 
 i\lambda{\rm \bf Tr} R_{\le j}  C^{1/2}_{\le j}  \vec \sigma   C_{\le j}\vec \sigma   C_{j} \vec \sigma C^{1/2}_{\le j} 
-\lambda^2 {\rm \bf Tr} R_{\le j}  D_{\le j}   C^{1/2}_{\le j}\vec \sigma   C_{j} \vec \sigma C^{1/2}_{\le j} \nonumber\\
&&\qquad\qquad +\ 
t{\rm \bf Tr}  C^{1/2}_{\le j}\vec \sigma   C_{j} \vec \Delta C^{1/2}_{\le j} \cdot  \left( \partial \vec \sigma ( -V_j )\right)   \biggr]
e^{- tV_{j}(\vec \sigma) }\ ,\nonumber
\eea
where in the last line we used integration by parts with respect to one $\sigma$ to explicit the cancellation in the last term.
In the last term the dot means a scalar product between the $\Delta$ and insertions of both the trace and the vertex derivative.
% and the notation $C_{\le j}$, $C'_{\le j}$ means $C_{\le j} (t_j)$, $C_{\le j} (t'_j)$. 
As expected this formula shows that the vacuum expectation value of the graph made of a single vertex has been successfully canceled by the counter term.
The contribution of a single vertex corresponds therefore again to perturbatively convergent graphs
with either at least two vertices, or one vertex and an operator $D$, multiplied by the exponential of the interaction, and can be treated therefore exactly as the ones with two or more vertices.

In all cases (including the single isolated blocks treated in \eqref{singbloc})
we apply a Cauchy-Schwarz inequality with respect to the positive measure $d\nu_\cB$ to separate the perturbative part ``down from the exponential" from the 
non-perturbative factor:
\bea  \label{CS} \vert  \int d \nu_\cB  F_\cB  \vert \le 
&& \sum_G  \Bigl( \int d \nu_\cB  \prod_a  e^{2 \vert V_{j_a} (\sigma_a) \vert }   \Bigr)^{1/2} 
\Bigl( \int d \nu_\cB   \vert  A_G (\sigma)  \vert^2  \Bigr)^{1/2} .
\eea

\subsection{Non-Perturbative Bound}

\begin{lemma}\label{boundv}
For $g$ in the cardioid domain $\cC  ard_\rho$
we have
\bea \vert V_j (\sigma) \vert &\le& \rho \;  O(1)   \bigl[ 1 +  {\rm \bf Tr}  \bigl(  C_{\le j}  \vec \sigma C_j   \vec \sigma \bigr)  \bigr]. \label{boundlemmanopert}
\eea
\end{lemma}
\prf  
Starting from \eqref{interaj4c}-\eqref{defDj} let us write $V_j = \cV_j + \cD_j$. Using the bound \eqref{boundcoun} for the $\cE'_j$ term, we get
\bea  \vert \cV_j (\sigma) \vert &\le &  \vert g \vert  \biggl( O(1) + 2 \int_0^1  t_j dt_j 
\vert {\rm \bf Tr}  \, R_{\le j}  C^{1/2}_{\le j}\vec \sigma   C_{j} \vec \sigma C^{1/2}_{\le j}    \vert   \biggr) .
\label{noper0}
\eea
For $A$ positive\footnote{We usually simply say positive for ``non-negative", i. e. each eigenvalue is strictly 
positive or zero.} Hermitian and $B$ bounded we have $\vert  \Tr A B \vert \le \Vert B \Vert \Tr A $. 
Indeed if $B$ is diagonalizable with eigenvalues $\mu_i$, computing the trace in a 
diagonalizing basis we have $\vert \sum_i   A_{ii} \mu_i \vert \le \max_i \vert \mu_i \vert \sum_i   A_{ii}  $;
if $B$ is not diagonalizable we can use a limit argument. 
Hence using \eqref{rescardbou}
\bee
{\rm \bf Tr} \vert  R_{\le j} C_{\le j} \vec  \sigma    C_j \vec \sigma \vert 
\le 2 \cos^{-1} (\phi /2)  {\rm \bf Tr}  \bigl(  C_{\le j}^{1/2} \vec \sigma C_j   \vec \sigma C_{\le j}^{1/2}  \bigr) = 2\cos^{-1} (\phi /2)
{\rm \bf Tr}  \bigl(  C_{\le j}  \vec \sigma C_j   \vec \sigma \bigr) . \label{noper1} 
\ee
We conclude that $\cV_j$ obeys the bound \eqref{boundlemmanopert} since in the cardioid $\vert g\vert \cos^{-1} (\phi /2) \le \rho$. 
It remains to check it for the $\cD_j$ term. Returning to \eqref{defDj}
\bee \vert \cD_j \vert  \le  \vert g \vert \int_0^1  dt_j 
 \vert {\rm \bf Tr} \, \vert g \vert^{1/2} R_{\le j} \bigl( D_{\le j}  C^{1/2}_{j} \vec \sigma C^{1/2}_{\le j}  +  C^{1/2}_{\le j}\vec \sigma C^{1/2}_{j} D_{\le j} +  C^{1/2}_{\le j}  \vec \sigma C^{1/2}_{\le j}D_j  \bigr)  \vert 
+  \vert g \vert  \vert {\rm \bf Tr}  R_{\le j}  D_{\le j}   D_j \vert . \label{Djboundb}
\ee
We use the Hilbert-Schmidt bound $ \vert {\rm \bf Tr} AB \vert \le {\rm \bf Tr} AA^\star + {\rm \bf Tr} BB^\star$. Remember \eqref{squaretraceclass}: $D$ is Hermitian positive and square trace class 
and so are also $D_j $ and $D_{\le j}$. Hence 
\bee   \vert {\rm \bf Tr}  R_{\le j}  D_{\le j}   D_j \vert  \le {\rm \bf Tr}  R^\star_{\le j}  R_{\le j}  D_{\le j}^2   +  {\rm \bf Tr}    D_j^2 \le O(1) [1+\cos^{-2} (\phi /2)] .
\ee
Similarly 
\bea  \vert {\rm \bf Tr} \, \vert g \vert^{1/2}  R_{\le j} D_{\le j}  C^{1/2}_{j} \vec \sigma C^{1/2}_{\le j} \vert &\le&  \vert g \vert  {\rm \bf Tr} R^\star_{\le j}  R_{\le j} D^2_{\le j} +   {\rm \bf Tr}   C_{\le j} \vec \sigma C_{j}  \vec \sigma, \nonumber\\
\vert {\rm \bf Tr} \, \vert g \vert^{1/2} D_{\le j} R_{\le j}  C^{1/2}_{\le j} \vec \sigma C^{1/2}_{j}  \vert &\le&  \vert g \vert{\rm \bf Tr}  D^2_{\le j}  R^\star_{\le j}  R_{\le j} +   {\rm \bf Tr}   C_{\le j} \vec \sigma C_{j}  \vec \sigma .
\eea
Finally for the last term $\vert {\rm \bf Tr} \, \vert g \vert^{1/2}   R_{\le j} C^{1/2}_{\le j} \vec \sigma C^{1/2}_{\le j} D_{j} \vert$, we remark that $ C^{1/2}_{\le j} D_{j} = t_j C^{1/2}_{j} D_{j}$. Then
\bee
t_j \vert {\rm \bf Tr} \, \vert g \vert^{1/2} D_{j}   R_{\le j} C^{1/2}_{\le j} \vec \sigma C^{1/2}_{j} \vert \le   \vert g \vert {\rm \bf Tr} R^\star_{\le j}  R_{\le j} D^2_{j} +   {\rm \bf Tr}   C_{\le j} \vec \sigma C_{j}  \vec \sigma .
\ee
Using again the inequality $\vert  \Tr A B \vert \le \Vert B \Vert \Tr A $ for $A$ positive and $B$ bounded, we can get rid of the resolvents:
\bee \vert g \vert  {\rm \bf Tr} R^\star_{\le j}  R_{\le j} D^2_{\le j} \le O(1)\vert g \vert \cos^{-2} (\phi /2) , \quad  \vert g \vert  {\rm \bf Tr} R^\star_{\le j}  R_{\le j} D^2_{j} \le O(1)\vert g \vert \cos^{-2} (\phi /2).
\ee
Hence we can conclude that the three first terms in \eqref{Djboundb} obey the bound \eqref{boundlemmanopert}  since in the cardioid $\vert g\vert \cos^{-1} (\phi /2) \le \rho$. \qed

\medskip
We can now bound the first factor in the Cauchy-Schwarz inequality \eqref{CS}.
\begin{theorem}[Bosonic Integration]\label{BosonicIntegration}
For $\rho $ small enough and for any value of the $w$ interpolating parameters
\bea  \Bigl{(} \int d\nu_\cB \prod_{a \in \cB}   e^{ 2 \vert V_{j_a} (\sigma_a) \vert }   \Bigr{)}^{1/2}  &\le& 
e^{ O(1) \rho  \vert \cB \vert }.
\eea
\end{theorem}
\prf  The term $\prod_{a \in \cB} e^{c \rho}$ is simply $e^{ c \rho  \vert \cB \vert }$. Applying Lemma \ref{boundv} we get
\bee
 \int d\nu_\cB  \prod_{a \in \cB}   e^{ 2 \vert V_{j_a} (\sigma^a)\vert  }  \le e^{ c \rho  \vert \cB \vert }
\int d\nu_\cB  \; e^{ \; < \sigma , \bQ \sigma > }
\ee
where $\bQ$ is a symmetric positive matrix in the big vector space  $\bV$ which includes color, components and vertices  indices $a \in \cB$.
This big space has dimension $N_\bV = 3 \vert \cB \vert N_{max}^2$. 
Hence the matrix $\bQ$ is an $N_\bV$ by $N_\bV$ matrix.
More precisely $\bQ$ is defined by the equation:
\bee  < \sigma , \bQ \sigma >  = \sum_{a \in \cB}  < \sigma^a , Q^a \sigma^a >, \  < \sigma^a , Q^a \sigma^a > \equiv
2\rho  \int_0^1 dt_{j_a} {\rm \bf Tr}  \bigl(  C_{\le j_a}  \vec \sigma^a C_{j_a}   \vec \sigma^a \bigr).
\ee
Hence $\bQ = \sum_{a \in \cB} \bQ^a $,
where $\bQ^a $ is the $N_\bV$ by $N_\bV$ matrix with all elements zero except the $3 N_{max}^2 $ by $3  N_{max}^2$
which have both vertex indices equal to $a$. These non zero elements form the
$3 N_{max}^2 $ by $3  N_{max}^2$ positive symmetric matrix $Q^a$ with matrix elements
\bea
Q^a_{c,m,n;\; c' m' n'} &=&  Q^{a,1}_{c,m,n;\; c' m' n'} + Q^{a,2}_{c,m,n;\; c' m' n'}\nonumber\\
Q^{a,1}_{c,m,n;\; c' m' n'}&=& 
 2\rho \delta_{c,c'} \int_0^1 dt_{j_a} 
 \delta_{m,m'} \delta_{n,n'}  [ t_{j_a}   \cQ_{j_a,j_a,1} (m,n)  + \sum_{k=0}^{j_a -1} \cQ_{j_a,k,1} (m,n) ]      \nonumber\\
 Q^{a,2}_{c,m,n;\; c' m' n'}&=& 
 2\rho (1-  \delta_{c,c'})  \int_0^1 dt_{j_a} \delta_{m,n}\delta_{m',n'}   [ t_{j_a}  \cQ_{j_a,j_a, 2} (m,m') + \sum_{k=0}^{j_a -1}  \cQ_{j_a,k, 2} (m,m')  ]  
 \eea 
where the $\cQ$ factors are defined respectively as the color-diagonal and color off-diagonal part of a bubble with 
two propagators of slices $j $ and $k$:
\bea
\cQ_{j,k,1} (m,n) &\equiv&   \sum_{m_2, m_3}  C_{k} (m, m_2, m_3 ) C_{ j} (n, m_2, m_3) ],
\label{boundd1}\\
\cQ_{j,k, 2} (m,m') &\equiv&   \sum_{m_3}  C_{k} (m, m', m_3 ) C_{ j} (m, m', m_3) \label{boundd2}.
\eea
The big matrix $\bQ$ has elements
$\bQ_{a,c,m,n;\; a',c' m' n'} =  \delta_{a,a'}  Q^a_{c,m,n;\; c' m' n'} $. 
Using the bounds \eqref{propmombound} it is easy to check that
\bea
\cQ_{j,k,1} (m,n) &\le& O(1)  M^{-2j}e^{- M^{-j} \vert n \vert }e^{- M^{-k} \vert m \vert }, \nonumber\\
\cQ_{j,k , 2} (m,m') &\le& O(1)  M^{-2j-k} e^{- M^{-k} (\vert m \vert + \vert m' \vert )} .
\eea

\begin{lemma}
The following bounds hold uniformly in $j_{max}$ and $N_{max}$
\bea
{\rm \bf Tr}\; Q^a &\le&  O(1)  \rho ,\label{bonnetrace}\\
\Vert Q^{a} \Vert  &\le& O(1) \rho  j_a  M^{-2 j_a} \label{bonnenorm}.
\eea
\end{lemma}
\prf  The first bound is easy. Since we compute a trace, only $Q^{a,1}$ contributes and the bound follows from \eqref{boundd1}
which implies that $\sum_{m, n} \cQ_{j,1} (m,n) \le O(1) $.
Since $Q^{a,1}$ is diagonal both in component and color space, from \eqref{boundd1}
we deduce that $\sup_{m, n} \cQ_{j,1} (m,n) \le O(1) j M^{-2j}  $, hence
\bee
\Vert Q^{a,1} \Vert  \le O(1)  \rho  j_a  M^{-2 j_a} \label{bonnenorm1}.
\ee
Finally to bound  $\Vert Q^{a,2} \Vert $
we use first a triangular inequality to sum over the 6 pairs of colors $c, c'$ and over $k$
\bee
\Vert Q^{a,2} \Vert  \le 12\rho \sum_{k=0}^{j} \Vert E_{j_a,k,2}\Vert    
\ee
where $ E_{j_a,k,2} $ is the (component space) matrix with matrix elements 
\bee E_{j_a,k,2} (m,n ; m',n')=\delta_{m,n}\delta_{m',n'}  \cQ_{j_a,k,2} (m,m') .
\ee
The operator norm of $E_{j_a,k,2} $ is bounded by its Hilbert Schmidt norm
\bee  \Vert E_{j_a,k,2}\Vert_2 = [ \sum_{m,m'} \cQ^2_{j_a,k,2} (m,m')]^{1/2}  \le   O(1) M^{-2j_a - k} [M^{2k}]^{1/2} = O(1) M^{-2j_a} .
\ee
It follows that
\bee
\Vert Q^{a,2} \Vert   \le O(1) \rho  j_a  M^{-2 j_a}, \label{bonnenorm2}
\ee
and gathering \eqref{bonnenorm1} and \eqref {bonnenorm2} proves \eqref{bonnenorm}.
\qed

The covariance $\bX$ of the Gaussian measure $d\nu_\cB$ is also a symmetric matrix on the big space $\bV$, but which is the tensor product 
of the identity in color and component space times the matrix $X_{ab} (w_{\ell_B} )$ in the vertex space. 
Defining $\bA \equiv \bX \bQ$, we have
\begin{lemma}
The following bounds hold uniformly in $j_{max}$ and $N_{max}$
\bea
\Tr \; \bA &\le&  O(1)  \rho  \, \vert \cB \vert  ,\label{bonnegtrace}\\
\Vert \bA \Vert  &\le& O(1) \rho  \label{bonnegnorm}.
\eea
\end{lemma}
\prf Since $\bQ = \sum_{a \in \cB} \bQ^a $ we find that 
\bee
\Tr \; \bA = \sum_{a \in \cB} \Tr \bX \bQ^a = \sum_{a \in \cB}  X_{aa} (w_{\ell_B} ) \Tr  Q^a = \sum_{a \in \cB} {\rm \bf Tr}\; Q^a \le  O(1)  \rho  \, \vert \cB \vert .
\ee
where in the last inequality we used \eqref{bonnetrace}. Furthermore by the triangular inequality and \eqref{bonnenorm}
\bea
\Vert \bA \Vert  \le   \sum_{a \in \cB} \Vert \bX \bQ^a \Vert = \sum_{a \in \cB}  X_{aa} (w_{\ell_B} ) \Vert  Q^a \Vert = \sum_{a \in \cB}  \Vert  Q^a \Vert 
\le  \sum_{j =0}^{\infty} 
O(1)  \rho  jM^{-2j} \le O(1).
\eea
where we used the fundamental fact that all vertices $a\in\cB$ have \emph{different scales} $j_a$.
\qed

We can now complete the proof of Theorem \ref{BosonicIntegration}. 
By \eqref{bonnegnorm} for $\rho$ small enough the series $\sum_{n=1}^\infty (\Tr \bA^n)/n $ converges and we have
\bea \int d\nu_\cB  \; e^{ \; < \sigma , \bQ \sigma > } &=& [\det (1  - \bA  )]^{-1/2}  = e^{-(1/2)\Tr \log  (1  - \bA  )  } = 
e^{(1/2) \sum_{n=1}^\infty (\Tr \bA^n)/n  } \nonumber
\\
&\le& e^{(1/2) \Tr \bA  (\sum_{n=1}^\infty \Vert \bA \Vert ^{n-1})  } 
\le e^{ O(1)  \rho  \vert \cB \vert}. 
\label{goodtra}
\eea
\qed

\subsection{Graph Bounds}

We still have to bound the second factor in \eqref{CS}, namely
$\Bigl( \int d \nu_\cB   \vert  A_G (\sigma)  \vert^2  \Bigr)^{1/2}$.
We recall that at fixed $|\cB|$, the graphs $G$ are forests with $E(G)= |\cB|-1$ (colored) edges joining 
$V(G) = c(G) + E(G)$ (effective) vertices, each of which has a weight given by 
\eqref{developcycle} and \eqref{developcycles}. The number of connected components $c(G)$ is bounded by $ |\cB|-1$, hence
\eqref{foreboun} holds.

This squared amplitude can be represented as the square root of an ordinary amplitude but for a graph $G"= G \cup G'$ 
which is the (disjoint) union of the graph $G$ and its mirror conjugate graph $G'$ of identical structure 
but on which each operator has been replaced by its Hermitian conjugate. This overall graph $G"$  
has thus twice as many vertices, edges, resolvents,  $\vec\sigma^a$ insertions and connected components than the initial graph $G$.

To evaluate the amplitude $A_{G"}= \int d \nu_\cB \vert  A_G (\sigma)  \vert^2$, we first delete every $\vec\sigma^a$ insertion using 
repeatedly integration by parts
\bea
\int \sigma^c_{n^c} F(\vec \sigma) d\nu (\sigma)=\int  \frac{\partial}{\partial \sigma^c_{n^c}}F(\vec \sigma) d\nu (\sigma) .
\eea
The derivatives $\frac{\partial}{\partial \sigma^c_{n^c}}$ will act on any resolvent $R_{ j_a}$ or remaining $\vec \sigma^a$ insertion of $G"$,
creating a new contraction edge. When it acts on a resolvent, it creates a new corner, bearing a $C^{1/2}_{\le j_a} R_{\le j_a} C^{1/2}_{\le j_a}$ (or $C^{1/2}_{\le j_a}R_{\le j_a}^{\dagger} C^{1/2}_{\le j_a}$) product of operators.

Remark that at the end of this process we have a sum over new graphs $\mathfrak G$ with no longer any $\vec\sigma^a$ insertion, but the
number of edges, resolvents and connected components at the end of this contraction process typically has changed. However we have a bound
on the number of new edges generated by the contraction process. Since each vertex of $G$ contains at most two $\vec\sigma^a$ insertions,
$G"$ contains at most $4 V(G)$, hence using \eqref{foreboun} at most $8(|\cB|-1)$ insertions to contract. Each such contraction creates 
at most one new edge. Therefore each graph $\mathfrak G$ contains  the initial $2(|\cB|-1)$ colored edges of $G"$ decorated with up to at most $8(|\cB|-1)$ additional new edges.

Until now, the amplitude $A(\mathfrak G)$ contains $C_{\le j}^{1/2}=\sum_{j'< j}C^{1/2}_{j'}+t_j C^{1/2}_j$ operators. We now develop the product of all such $C^{1/2}_{\le j}$ factors as a sum over scale assignments
$\mu$, as in \cite{Riv}. It means that each former $C^{1/2}_{\le j}$ is replaced by a fixed scale $C^{1/2}_{j'}$ operator (the $t_j$ factor being bounded by 1)
with  scale attribution $j'\leq j$. The amplitude
at fixed scale attribution $\mu$ is noted $A(\mathfrak G_\mu)$ and we shall now bound each such amplitude. The sum over $\mu$ will be standard to bound
after the key estimate \eqref{decayattri} is established. Similarly the sums over $G$ and over $\mathfrak G$ only generate
a finite power of $\vert \cB \vert !$, hence will be no problem using the huge
decay factors of  \eqref{decayattri}.

\begin{theorem}[Graph bound]\label{thmgraphbound}
The amplitude of a graph $\mathfrak G$ with scale attribution $\mu$ is bounded by
\bea
|A(\mathfrak G_\mu)| &\leq& [O(1) \rho]^{E(\mathfrak G)}
M^{-\frac12\sum_{v\in G} j_{a(v)}}.
\eea
\end{theorem}
 
\prf
We work at fixed value of each $\sigma$, and denote $\mathfrak C$ the connected components of a graph $\mathfrak G$, thus $A(\mathfrak G_\mu)=\prod_{\mathfrak C}A(\mathfrak C_\mu)$.
The amplitude $A(\mathfrak C_\mu)$ of a connected component $\mathfrak C$ can be bounded by iterated Cauchy-Schwarz inequalities \cite{sefu1,Delepouve:2014bma} using the formula
\begin{align}\label{cauchy} \vert
\bra{\alpha} R\otimes R'\otimes \mathbf{1}^{\otimes p}\ket \beta \vert \;   \leq  \;  \|R\|\|R'\|\sqrt{\braket {\alpha|\alpha}}\sqrt{\braket{\beta|\beta}} \; ,
\end{align}
where the scalar product $\braket {f | g}$ means the scalar product in the natural three stranded Hilbert space $\cH$.

First, we choose a spanning tree for each connected component $\mathfrak C$, and order the resolvents $R_{\le j}$ at the corners along the clockwise 
contour walk of the tree. For simplicity we consider first a connected component with an even number $2n$ of resolvents, which are then labeled from $R_1$ to $R_{2n}$.

We choose $R_1$ and the antipodal resolvent $R_{n+1}$ as the marked operators $R$ and $R'$ to apply \eqref{cauchy}. Hence we split the tree in two parts, 
according to the unique path going from the corner 1 to the corner $n$. The vector $\alpha$ is made of everything on the left on the splitting line, 
and the vector $\beta$ is made on everything on the right. The identity  $\mathbf{1}^{\otimes p}$ comes from all loop edges which cross the splitting line, as in \cite{Delepouve:2014bma}.

The graphs $\braket {\alpha|\alpha}$ and $\braket{\beta|\beta}$ have the same structure of plane trees decorated with loop edges and a product of operators on the corners,
but each one has only $(2n-2)$ resolvents left. 
%Each graph now has two ``cleaned" corners, which bear two (identical) $C^{1/2}_{j'}$ but no resolvent (and eventually some $D$ operators).
We can repeat the same cleaning process on each new graph by ordering the $(2n-2)$ resolvents along the clockwise contour walk of the new graph,
then choosing a new pair of antipodal resolvents as $R$ and $R'$. 
Repeating the process $n$ times gives thus a geometric mean over $2^n$ final completely cleaned graphs ${\mathfrak g}$ bearing no $R_{\le j}$ at all, 
times a product of norms of resolvents, which are all bounded by $2\cos^{-1}(\phi/2)$ for $g=|g|e^{i\phi}$ in the cardioid. Since these graphs no longer have
any dependence on $\sigma$, the normalized measure $\int d\nu ( \sigma)$ simply evaluates to 1, and we are left with a \emph{perturbative} bound:
\bea
|A({\mathfrak C_\mu})| &\leq&  \prod_{i=1}^{2n}\|R_i\|\ \left( \prod_{\mathfrak g} A({\mathfrak g}) \right)^{\frac1{2^n}}
 \leq 4^n \left[\cos\frac\phi2\right]^{-2n} \left( \prod_{\mathfrak g} A({\mathfrak g}) \right)^{\frac1{2^n}}.
\eea
The Cauchy-Schwarz process keeps track of a number of items. Indeed, at every iteration, 
each vertex and edge of the bounded graph gives respectively two vertices and two edges of the next-stage graphs.
The $C^{1/2}_j$ and $A_j$ operators follow the same rule, and each $C_j^{1/2}$ operator of the original graph will generate $2^n$ identical
$C_j^{1/2}$ operators in the final graphs, which repartition is {\it a priori} unknown. 
Finally, each vertex of the original graph bearing at least one resolvent, each vertex of the final graph has been cut at least once and is thus mirror-symmetric.
 
In a final graph $\mathfrak{g}$, each corner bears either a $C_j$ operator or a product of two identical $C_{j'}^{1/2}$ operators (thus one full $C_{j'}$), 
vertices may also bear $A_j$ insertions, and the strands represent contractions of their indices. All operators left are diagonal, and bounded as 
%the $A_j$ operators are bounded  by $O(1){\rm log} j$and the $C^{1/2}$ by
\bea
C_j^{1/2}(n, \bar n)&=& \delta_{n,\bar n}  \sqrt{\frac{1}{n^2+1}} ({\bf 1}_{M^{2j-2}<n^2+1 \le M^{2j}})\le \frac{1}{M^{j-1}}\delta_{n,\bar n}
\prod_{i=1}^3{\bf 1}_{n_i^2 \le M^{2j}}, \nonumber\\
A_j(n, \bar n)&=& {\bf 1}_j(n)\delta_{n\bar n} \sum_c A(n_c)  \le \delta_{n\bar n}\sum_c O(1) {\rm log} M^j =j O(1)\delta_{n\bar n}
\ .
% %\int_{M^{-2j}}^{M^{-2(j-1)}}e^{-\alpha (n_1^2+n_2^2+n_3^2 + 1)} \ d\alpha\nonumber\\
% &=&  \int_{M^{-2j}}^{M^{-2(j-1)}}d\alpha\  e^{-\alpha} \prod_{i=1}^3  e^{-\alpha n_i^2}\delta_{n_i,\bar n_i}  \nonumber \\ 
% & \le &  \ O(1) M^{-2j} \prod_i e^{-M^{-j} \vert n_i \vert    }\delta_{n_i,\bar n_i} \;\; {\rm for}\; j\ge 1.
\eea
 Then, for a final graph $\mathfrak{g}$ with $2n$ corners that we index by $\eta$, each bearing a $C^\eta$ operator, and denoting $a(\eta)$ the vertex of the original graph that bore the operator,
\bea
A(\mathfrak g) &\leq& |\lambda|^{2E(\mathfrak g)}\sum_{\{\vec n\}} \prod_\eta C^{\eta}(n_{\eta} \bar n_{\eta} )\delta_{\bar n_\eta n_{\eta}}\ \prod_{{\rm strands}\ s} \delta_{n_{i_s}^s \bar n_{i_s}^s}\prod_{{\rm operators}\ A}O(1)j_{a(A)} \nonumber\\
&\leq&|\lambda|^{2E(\mathfrak g)}\sum_{\{\vec n\}}\  \prod_\eta \delta_{n_\eta,\bar n_{\eta}}
\frac{1}{M^{2j_\eta-2}}\left(\prod_i {\bf 1}_{n_{i\, \eta} \le M^{j_\eta}}\right)\  
\prod_{ s} \delta_{n_{i_s}^s \bar n_{i_s}^s} \prod_{A}O(1)j_{a(A)}\\
&=& [M^2 |\lambda|^2]^{E(\mathfrak g)}\ M^{-2\sum_\eta j_\eta}\ \prod_{{\rm faces}\ f}\sum_{ n_f}\prod_{\eta, \eta'\in f}\left( {\bf 1}_{n_f \le M^{j_{\eta}}}\right)\prod_{A}O(1)j_{a(A)} ,
%e^{-  (\sum_{\eta\in f}  M^{-j_\eta}) \vert n_{f} \vert    }\ ,\nonumber
\nonumber\eea
where $j_\eta \in \{0...j_{a(\eta)}\}$ is the scale assignment of the corresponding $C$ operator, $j_{a(A)}$ the scale of the vertex bearing the operator $A_{j}$ and $f$ are the faces of color $i$. 
In the bound, the $C$ operators being removed, the 
faces are closed cycles of $\delta$ operators multiplied by scale factors and cutoffs. Hence only one index $n_f$ remains for each colored face $f$. 
Thus the amplitude of a final graph $\mathfrak g$ is bounded by
\bea
A(\mathfrak g) &\leq&  [M^2 |\lambda|^2]^{E(\mathfrak g)} M^{-2\sum_\eta j_\eta}\ \prod_{f} M^{j_{min}(f)} \prod_{A}O(1)j_{a(A)}\nonumber\\
&\leq&  [O(1) |g|]^{E(\mathfrak g)} M^{\sum_f j_{min}(f)-2\sum_\eta j_\eta} \prod_{A}O(1)j_{a(A)}.
\eea
Corners of final graphs that were generated by distinguished corners (without resolvents) of the original graph will be denoted $\eta^* \in H^*$, 
as opposed to regular corners $\eta$. Those corners bear $C_{j_{a(\eta^*)}}$ operators that we want to keep track of. Other corners bear $C$ operators of scale $j_\eta\leq j_{a(\eta)}$.
The amplitude is thus bounded by
\bea
|A({\mathfrak C_\mu})| &\leq&\left[\frac12 \cos\frac\phi2\right]^{-2n} \left( \prod_{\mathfrak g}  [O(1) |g|]^{E(\mathfrak g)} \prod_{A}O(1)j_{a(A)}\right)^{\frac1{2^n}}
M^{\frac1{2^n}\sum_{\mathfrak g}\left[\sum_f j_{min}(f)-2 j_\eta \right]}\nonumber\\
&\leq&   [O(1) \rho]^{E(\mathfrak C)}\left(\prod_{v}j_{a(v)}\right)
M^{-\frac12\sum_v j_{a(v)}}\ M^{\frac1{2^n}\sum_{\mathfrak g}\left[\sum_f j_{min}(f)-2\sum_{\eta} j_\eta +\frac{1}{2}\sum_{H^*}j_{\eta*}\right]}\nonumber\\
&\leq&   [O(1) \rho]^{E(\mathfrak C)}
M^{-\frac14\sum_v j_{a(v)}}\ M^{\frac1{2^n}\sum_{\mathfrak g}\left[\sum_f j_{min}(f)-2\sum_{\eta} j_\eta +\frac{1}{2}\sum_{H^*}j_{\eta*}\right]},
\eea
where we use the conservation of the number of distinguished corners during the Cauchy-Schwarz process $\sum_{\mathfrak g}\left[\sum_{H^*}j_{a(\eta*)}\right]=
2^n \sum_v j_{a(v)}$, along with the fact that for any graph, $2n< 2E(\mathfrak C)$. We also used the conservation of $A_j$ operators, and the fact that there is at most one $A_j$ per original vertex.

For a connected component with an odd number $2n+1$ of resolvents, we first proceed to a slightly asymmetric
Cauchy-Schwarz splitting of the graph, choosing $R^1$ and $R^{n+1}$ as $R$ and $R'$. Both scalar product graphs will then have an even number
of resolvents and the previous results stand.

\begin{lemma} 
For any connected components $\mathfrak C_\mu$ with final graphs $\mathfrak g$,
 \bea
\sum_{\mathfrak g}\left[\sum_f j_{min}(f)-2\sum_{\eta} j_\eta +\frac{1}{2}\sum_{H^*}j_{\eta*}\right]\leq0.
\eea
\end{lemma}

\prf
A final graph consists of the gluing of two mirror symmetric graphs along a path whose ends are undistinguished corners $\eta\not\in H^*$. 
Thus a final graph bears at least two undistinguished corners. Therefore,
\bea
\sum_f 1-2\sum_{\eta} 1 +\frac{1}{2}\sum_{H^*}1 = F - 2C + \frac12|H^*| \leq F-\frac32C -1.
\eea
For any tree, the relationship between the number $C$ of corners $\eta$, the number $F$ of faces $f$ and the number of $A_j$ insertions $|A|$ is $F-C+|A|=3$. 
This can be proved starting with a single isolated vertex and adding extra vertices, edges and $A_j$'s one by one. 
Each new vertex and edge comes with two new faces and two new corners, each $A_j$ with one corner, and the isolated vertex had three faces and no corner.

Any loop edge adds two corners and may increase or decrease the number of faces by one. 
Thus, for a tree $\mathcal T$ decorated with $L$ loop edges $\ell\in\mathcal L$, 
\bee (F-\frac32C-1)_{\mathcal T+\mathcal L} \leq (F-\frac32C-1)_{\mathcal T}-2L=3-V-\frac{3}{2}|A|-2L. \ee 
For any graph with at least $3$ vertices, or with at least one loop edge, or with $A_j$ insertions ($|A|$ is always even for a final graph), this is lower than $0$. 
A pathological final graph cannot be a single vertex without edges, because final graphs have at least two corners. 
A final graph composed of two vertices, no loops and no $A_j$ can only arise from a graph which had two consecutive corners bearing resolvents, 
separated by an edge of the chosen tree, before the last iteration of the Cauchy-Schwarz process.

\begin{figure}[!h]
  \begin{center}
  {\includegraphics[width=0.4\textwidth]{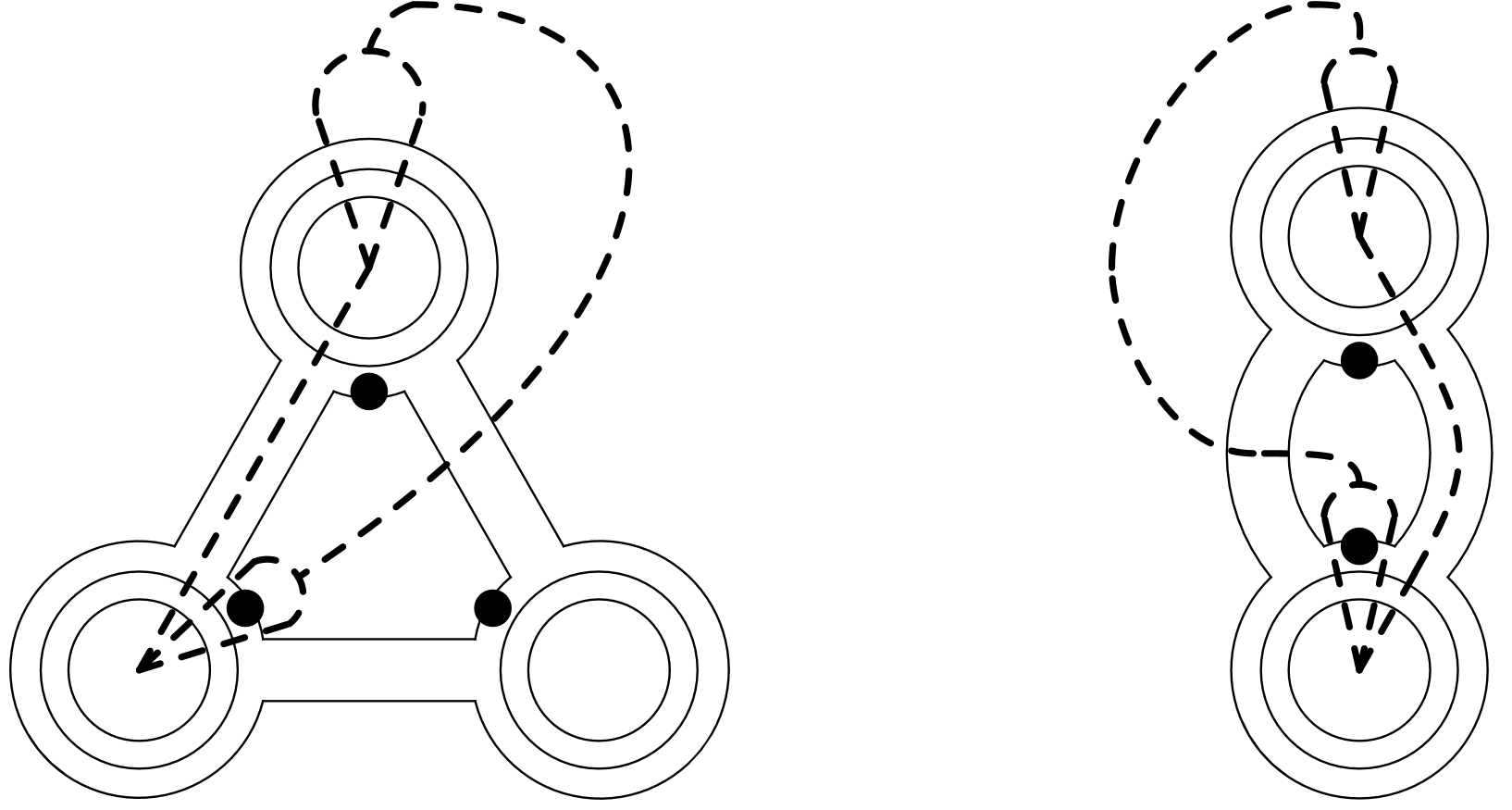}}
   \end{center}
  \caption{The graphs $S_3$ (left) and $S_2$ (right) with dashed lines representing the Cauchy-Schwarz splitting used to avoid pathological graphs.
  Dotted corners bears resolvents $R_{\le j}$. When the dashed line crosses an un-dotted corner, propagator $C_j$ must be rewritten as $C^{1/2}{\mathbf 1}C^{1/2}$ and the identity matrix $\mathbf{1}$ is used instead of a resolvent.}
  \label{S3S2}
\end{figure}

If a mirror-symmetric graph has only two resolvents, then those resolvents are mirror symmetric and therefore are on 
each side of the symmetry axis, which is a path between two ``cleaned'' corners (bearing no resolvent), therefore there is at least one corner between them. 
Therefore, a graph with only two remaining resolvents, which are on consecutive corners, cannot arise from the bounding process. 
There are only two families of original graphs with less than four resolvents, two being separated only by an edge. 
We will call them $S_2$ and $S_3$, and deal with them with an adapted Cauchy-Schwarz bound that avoid pathological final graphs (Fig. \ref{S3S2}).

Therefore, for any final graph, the $j$s brought by corners (2 for undistinguished ones, and $3/2$ for distinguished ones)
is large enough to cancel the number of $j_{min}$ brought by the faces. However, each $j_{min}$ must be canceled individually by a higher $j$.

First, we consider a distinguished corner of scale $j_{a(\eta^*)}$. Such a corner is generated by a corner without resolvent and thus 
cannot be used in a Cauchy-Schwarz bound. Thus, each vertex being mirror symmetric, they carry an even number of distinguished corners.
If a vertex only bears distinguished corners, it is then made of $2k$ replicas of the same corner (and thus brings $3kj_{a}$), and has degree $2k$. 
A vertex of degree $2k$ can belong to at most $2+2k$ faces. For $k>1$, the $3kj_a$ are enough to cancel the $j_{min}$ of all faces the vertex belongs to. 
For k=1 (vertex of degree 2), if the two edges are of different colors, the vertex belongs to only three faces, that are canceled out by the $3j_a$. 
If the two edges are of the same color $c$, the vertex can belong to two distinct faces of color $c$. 
If any of those faces also goes through a vertex of degree two bearing two undistinguished
corners (and thus bringing $4j_\eta$, enough to cancel all the faces running through it),
or a vertex of degree $\geq 2$, its $j_{min}$ will be canceled out by this vertex.
If both those faces run only through vertices of degree two bearing only distinguished corners, 
then the final graph must be a closed cycle of vertices of degree two bearing only distinguished corners, which is impossible.
Therefore, $\frac32\sum_{H^*}j_{\eta^*}$ is enough to cancel out every potential faces with $j_{min}=j_{a(\eta^*)}$.

For vertices bearing undistinguished corners, the situation is actually better. Each vertex of degree $>1$ brings enough $j_\eta$ to cancel each faces it belongs to. 
Only the leaf without $A_j$ has one more face than $j_\eta$s. However, one face running through a leaf will also run through its only neighboring vertex, 
which is of degree two or more (recall that the two-leaves-graph is excluded), and will be canceled out by this vertex. If the neighbor is of degree two, 
then it has only $3$ faces running through it. If it is of degree $>2$, then it has more than enough $j_\eta$.  

Therefore on any final graph, all the $j_{min}$ can be canceled individually by a $j_\eta$, hence we have
\bea
\sum_{\mathfrak g}\left[\sum_f j_{min}(f)-2\sum_{\eta} j_\eta +\frac{1}{2}\sum_{H^*}j_{\eta*}\right]\leq0,
\eea
and thus,
\bea
|A(\mathfrak G_\mu)| =\prod_{\mathfrak C_\mu} |A(\mathfrak C_\mu)|&\leq& [O(1)\rho]^{E(\mathfrak G)}
M^{-\frac14\sum_{v\in G} j_{a(v)}}.  \label{decayattri}
\eea\qed

\section{Conclusion}

Once decay in the maximal scale at each vertex has been garnered by \eqref{decayattri} 
the remaining sum over scale attributions $\mu$ is completely standard \cite{Riv}. Similarly 
the auxiliary sums such as those over $\tau$ and the other terms in \eqref{developcycles}, over partitions $\pi$ in \eqref{partitio} 
(hence over the choice of $G$) and over $\sigma$ contractions (hence over the choice of $\mathfrak G$)
cannot endanger convergence, exactly as in  \cite{Rivasseau:2014bya,Riv}. Indeed the key observation is that in a block $\cB$
\emph{since all slice indices are different} and since we cleaned first a distinguished propagator $C_{j_a}$
whose decay cannot have disappeared in  \eqref{decayattri}, any small power 
of the product $\prod_{j \in \cB} e^{- \frac12\sum_{v\in G} j_{a(v)}}$ is still smaller than $e^{- O(1) \vert \cB \vert ^2 }$
for some small $O(1)$, hence amply sufficient
to beat any fixed power of $\vert \cB \vert !$, such as those generated by the previous sums.

Combinatorial estimates are also exactly similar to those of \cite{MLVE} except for the fact
that counting the colored two-level trees requires an additional factor $3^{\vert \cF^{3c}_B\vert }$
to choose the colors of Bosonic edges. Hence
\begin{prop}\label{prop:counting}
 The number of two level trees with a three-colored first level over $n\ge 1$ vertices is bounded by $12^{n} n^{n-2}$.
\end{prop}

Uniform Taylor remainder estimates at order $p$ are required to complete the proof of Borel summability \cite{Sokal} in Theorem \ref{thetheorem}. 
They correspond to further Taylor expanding beyond trees up to graphs with \emph{excess} (ie number of cycles) at most $p$. The corresponding 
\emph{mixed expansion} is described in detail in \cite{Gurau:2013pca}. The main change is to force for an additional
$p!$ factor to bound the cycle edges combinatorics, as expected in the Taylor uniform remainders estimates of a Borel summable function.

The main theorem of this paper clearly also extends to cumulants of the theory,
introducing \emph{ciliated} trees and graphs as in \cite{Gurau:2013pca}. This
is left to the reader. Indeed in tensor theories the relation between such cumulants and
ciliated trees in the intermediate field representation is complicated, involving in the general case graphical branching 
of the cilia and Weingarten functions \cite{Gurau:2013pca,Delepouve:2014bma}, and could detract the reader's attention from what 
is new in the tensor field theory case. 

The next tasks in constructive tensor field theories would be to treat the $T^4_4$ and $T^4_5$, which correspond in 
level of difficulty respectively to $\phi^4_3$ and $\phi^4_4$ in the ordinary quantum field theory context with local interactions.
This would clearly require a much more precise phase space cell expansion. 
The reward is that ultimately, in contrast with $\phi^4_4$, a renormalizable tensor field theory 
such as $T^4_5$ should exist non-perturbatively without cutoffs, since it is asymptotically free \cite{BenGeloun:2012pu}.

\end{document}